\documentclass[12pt]{article}
\usepackage{amsmath,amsfonts,bm}
\usepackage{setspace}
\usepackage{natbib}
\usepackage{url}
\usepackage{algorithm}
\usepackage{algorithmic}
\usepackage{graphicx}
\usepackage{amsthm}
\usepackage{booktabs}

\newcommand{\sig}{\sigma}
\newcommand{\Sig}{\Sigma}

\newcommand{\cb}{\bm{z}}
\newcommand{\Gh}{\widehat{G}}
\newcommand{\Ghs}{\widehat{G}^\star}
\newcommand{\rhoh}{\widehat{\rho}}
\newcommand{\cbh}{\widehat{\cb}}
\newcommand{\Fh}{\widehat{F}}
\newcommand{\NN}{\mathcal{N}}
\newcommand{\EE}{\mathbb{E}}
\newcommand{\CDF}{\text{CDF}}
\newcommand{\GMM}{\text{GMM}}
\newcommand{\MW}{\text{MW}}
\renewcommand{\Re}{\mathbb{R}}
\newcommand{\PP}{\mathcal{P}}

\newcommand{\widgraph}[2]{\includegraphics[keepaspectratio,width=#1]{#2}}

\def\argmax{\textnormal{arg}\max}
\def\argmin{\textnormal{arg}\min}

\newtheorem{theorem}{Theorem}
\newtheorem{proposition}{Proposition}
\newtheorem{definition}{Definition}
\newtheorem{corollary}{Corollary}%opening

\newcommand{\blind}{0}

\usepackage{xcolor}[dvipsnames]

\def\stackover#1#2{\mathrel{\mathop{#2}\limits^{#1}}}
\newcommand{\iid}{\stackover{\mbox{\footnotesize i.i.d.}}{\sim}}

\newcommand{\partitle}[1]{} % change this to re-surrect some of the
\begin{document}

\def\spacingset#1{\renewcommand{\baselinestretch}%
{#1}\small\normalsize} \spacingset{1}

%%%%%%%%%%%%%%%%%%%%%%%%%%%%%%%%%%%%%%%%%%%%%%%%%%%%%%%%%%%%%%%%%%%%%%%%%%%%%%

\if0\blind
{
  \title{\bf Summarizing Bayesian Nonparametric Mixture Posterior - Sliced Optimal Transport Metrics for Gaussian Mixtures}
  \author{Khai Nguyen\\
    Department of Statistics and Data Sciences, University of Texas at Austin\\
    and \\
    Peter Mueller \\
    Department of Mathematics, University of Texas at Austin}
  \maketitle
} \fi

\if1\blind
{
  \bigskip
  \bigskip
  \bigskip
  \begin{center}
    {\LARGE\bf Title}
\end{center}
  \medskip
} \fi

\bigskip
\begin{abstract}
  Existing methods  to summarize posterior inference 
  for mixture models focus on identifying
a point estimate of the implied random partition for clustering, with
density estimation as a secondary
goal~\citep{wade2018bayesian,dahl2022search}. We propose a novel
approach for summarizing posterior  inference in 
nonparametric Bayesian mixture models, prioritizing estimation
of the mixing measure (or mixture) as an inference target. One of the
key features is the model-agnostic nature 
of the approach, which remains valid under arbitrarily complex
dependence structures in the underlying sampling model. Using a
decision-theoretic framework, our method identifies a point estimate
by minimizing posterior expected loss.
 A loss function is defined as a discrepancy
between mixing measures. Estimating the mixing
measure implies inference on the mixture density and the random partition. Exploiting the
discrete nature of the mixing measure, we use a version of sliced
Wasserstein distance. We introduce two specific variants
for Gaussian mixtures. The first, mixed sliced Wasserstein, applies
generalized geodesic projections on the product of the Euclidean space
and the manifold of symmetric positive definite matrices. The second,
sliced mixture Wasserstein, leverages the linearity of Gaussian
mixture measures for efficient projection\footnote{The code for the paper is published at \url{https://github.com/khainb/sbnpm-sot}.}.
% Empirical analyses on
% simulated and real data demonstrate that the proposed approach
% achieves comparable clustering performance while yielding more
% accurate density estimation. 
\end{abstract}

\noindent%
\vfill

\spacingset{1.75} % DON'T change the spacing!
\section{Introduction}
\label{sec:intro}
We propose an approach  to summarize  the posterior
distribution on the random mixing measure in Bayesian nonparametric
(BNP) mixture models~\citep{ghosal2017fundamentals}.
 By focusing on the mixing measure, 
the method fills a gap in
the existing literature which focuses on summarizing the posterior
distribution of implied random partitions of experimental
units~\citep{wade2018bayesian,dahl2022search}. The proposed approach
requires only posterior Monte Carlo samples of random mixing measures
as input and provides a well-defined summary of these
measures. {The approach is valid with any posterior simulation method.} Importantly, the approach is model-agnostic, accommodating
any prior distributions on the random mixing measures, including those
with dependent structures such as dependent Dirichlet
process~\citep{quintana2022dependent,maceachern:99}. Using the resulting
point estimate of the mixing measure, if desired, one can 
derive an estimate of the density function and the partition. 

 Most approaches to 
posterior summarization for BNP mixture models
 assume a context of density estimation, conditioning on a sample
\( \{Y_1, \ldots, Y_n\} \) from a mixture.
Latent indicators $z_i$,
$i=1,\ldots,n$ that link the data with atoms in the mixing measure
define a random partition $\rho=\{S_1,\ldots,S_K\}$ of
$[n] := \{1,\ldots,n\}$ into subsets $S_k$ of matching $z_i$. Methods
then aim to summarize $p(\rho \mid Y_1,\ldots, Y_n)$, or equivalently,
$p(\cb \mid Y_1,\ldots,Y_n)$ for $\cb=(z_1,\ldots,z_n)$.

% focuses on point estimates for the implied random
% partition. Specifically, given a dataset of \( n \) samples, \( \{Y_1,
% \ldots, Y_n\} \), these works aim to summarize \( p(\rho \mid Y_1,
% \ldots, Y_n) \) or \( p(\mathbf{c} \mid Y_1, \ldots, Y_n) \), where \(
% \rho = \{S_1, \ldots, S_k\} \) represents a partition of the indices
% \( \{1, \ldots, n\} \):=[n], and \( \mathbf{c} = (c_1, \ldots, c_n) \)
% denotes cluster membership indicators that take values in \( \{1,
% \ldots, k\} \).
Proceeding under a decision-theoretic framework involves minimizing
the posterior expectation of a chosen loss function to define the
Bayes rule: 
\begin{align}
    \label{eq:clustering_posterior_expected_loss}
\hat{\rho}^\star = \argmin_{\hat{\rho}} \mathbb{E}[\mathcal{L}(\rho, \hat{\rho}) \mid Y_1, \ldots, Y_n] \quad \text{or} \quad \hat{\cb}^\star = \argmin_{\hat{\cb}} \mathbb{E}[\mathcal{L}(\cb, \hat{\cb}) \mid Y_1, \ldots, Y_n],
\end{align}
where \( \mathcal{L} \) is a loss function that can be expressed
using both the partition or the cluster label
notation.
\citet{lau2007bayesian} propose the use of Binder
loss~\citep{binder1978bayesian}. \citet{wade2018bayesian} use
variational information (VI) loss~\citep{meilua2007comparing} as an
alternative. Other information-based distances include normalized
variational information (NVI), information distance (ID), normalized
information distance (NID)~\citep{vinh2009information},
one minus adjusted Rand index (omARI)~\citep{rand1971objective}, generalized Binder and VI~\citep{dahl2022search}. In
addition to selecting the loss function, various search algorithms
have been proposed to solve the optimization problem in
~\eqref{eq:clustering_posterior_expected_loss}, including binary
integer programming~\citep{lau2007bayesian}, greedy search based on
the Hasse diagram~\citep{wade2018bayesian}, the R \& F
algorithm~\citep{rastelli2018optimal}, and the SALSO
algorithm~\citep{dahl2022search}.

Even when the primary inference target is a posterior summary of
$\rho$,  estimation of the mixture or the mixing measure can
still be reported in a second step {via additional sampling. Also, in many cases it is convenient to focus  on the mixing measure since the mixing measure implies both the partition and the density. Moreover, summarizing the mixing measure can avoid dealing with label switching problems that arises when summarizing the partition.} 
% Mixture models are well-known for their effectiveness in solving
% clustering problems and performing density estimation.
% However,
% current Bayesian analysis methods prioritize summarizing the random
% partition for clustering. From this summarized partition, density
% estimation is inferred secondarily through sampling.
% However, in applications
% where the density function plays a crucial role, such as anomaly
% detection and data generation, it may be more appropriate to focus
% directly on the density function.
% In such cases, a point estimate of the partition may not be the
% primary objective. 
Consider then a general BNP mixture model: 
$
    Y_1,\ldots,Y_n \overset{i.i.d}{\sim} F, \quad F = f*G, \quad G \sim p(G),
$
where $f$ is a kernel, $*$ denotes the convolution operator, {$F$ is the mixture density, and $G$ is the mixing measure (or the mixture)}. We
propose to find a summary for $p(G|Y_1,\ldots,Y_n)$.
From a point estimate $\Gh$, one can
directly obtain a point estimate of the density $\Fh$ {as
a convolution} with the density kernel $f$.
 And if desired,  one can obtain a point estimate $\rhoh$
(or cluster labels $\cbh$) induced by $\Gh$.
For the sake of easy exposition, in the upcoming discussion we assume
i.i.d. sampling. But we do so without loss of generality. The only
assumption is that posterior inference is provided as a posterior
Monte Carlo sample of $G$. The details of the sampling model, prior
model, {and the details of the posterior Monte Carlo simulation scheme} can be arbitrary. 

{Similar to \eqref{eq:clustering_posterior_expected_loss}}, we continue to use a decision-theoretic approach by minimizing a
posterior expected loss. However,
{targeting} $G$, 
% in our case,
we require a loss function that quantifies discrepancy between two
measures.
% define the posterior expected loss.
Given the discrete nature of the random measure $G$, optimal transport
distances~\citep{villani2009optimal,peyre2020computational} are a
natural choice in this context. Specifically, we choose
sliced Wasserstein (SW)
distance~\citep{rabin2012wasserstein,bonneel2015sliced} due to its
computational and statistical scalability with respect to the number
of {dimensions and the number of}  support points. With  near-linear time complexity in the number of
support points, the SW distance facilitates efficient and accurate
truncation of mixing measures. 

{Recall that our primary inference aim is summarizing the mixing measure. If the primary goal were the mixture density, a more natural option would be to work with less parsimonious, but easier tractable location mixtures. Considering general mixture of Gaussians}, we introduce two novel
versions of SW, working with 
% sliced Wasserstein distances between mixing measures, which are
measures supported on the product of the Euclidean manifold and the
manifold of symmetric positive definite (SPD) matrices.  The first
variant, called mixed SW (Mix-SW), uses generalized
geodesic projection instead of the conventional linear projection used
in standard SW. Consequently, Mix-SW preserves more geometric
information compared to SW with a vectorization approach. The second
variant, named sliced mixture Wasserstein (SMix-W), compares mixing
measures by evaluating the induced mixture of Gaussian measures. By
leveraging the linearity properties of mixture of Gaussian measures,
SMix-W achieves a reduction in projection complexity compared to
traditional SW while still being geometrically meaningful. Finally, we
discuss basic properties of the proposed distances including
boundedness and metricity. 

The remainder of the article is organized as follows: In
Section~\ref{sec:point_estimation_mixing_measures}, we introduce the
approach for summarizing the posterior of random mixing
measures. Section~\ref{sec:SWGMM} presents the two novel distances for
Gaussian mixing measures, accompanied by a discussion of their
theoretical and computational properties. In
Section~\ref{sec:experiments}, we
conduct an empirical analysis using
simulated data and the Old Faithful Geyser dataset, employing a
truncated Dirichlet Process Gaussian mixture model. We assess
% the performance of the
the proposed approach in both clustering and density estimation.
% Finally, we draw conclusions in Section~\ref{sec:conc}.
Additional materials, including technical
proofs, are provided in the appendices. {We refer the reader to Appendix~\ref{subsec:acronyms} for the list of acronyms in the paper.}

\section{Point estimation of random mixing measures}
\label{sec:point_estimation_mixing_measures}
We present a new approach for obtaining a point
estimate of random mixing measures. % using decision theory.
Without loss of generality 
we consider the {setting of inference under} following BNP mixture model:  
$
    Y_1, \ldots, Y_n \iid F, \quad F = f * G, \quad G \sim p(G),
$
where $f$ is a kernel and $p(G)$ denotes a prior on the random mixing
measure $G$. Our objective is to report a point estimate
$\Gh$.
The point estimate can then be used for any downstream data
analysis. 
We first define the problem in
Section~\ref{subsec:problem_setup} and then discuss current options
for optimal transport distances that can be used as the loss function
in Section~\ref{subsec:otdistances}.

\subsection{Problem Setup}
\label{subsec:problem_setup}

\partitle{Posterior expected loss.} 
% \mynote{K: i supressed some of the
%   par titles. Think they are great! But not really common for a stats
%   paper. I kept some later, when it's most helpful}
  
{Let $\Theta$ denote the space of supports of the mixing measure $G$}. Under a decision-theoretic framework, the point estimate of the mixing
measure $\Ghs$ minimizes posterior expected loss: 
\begin{align}
\label{eq:posterior_expected_loss}
    \Ghs = \text{arg}\min_{\Gh} \mathbb{E}[\mathcal{D}(G, \Gh)
  \mid Y_1, \ldots, Y_n], 
\end{align}
where $\mathcal{D}: \PP(\Theta) \times \PP(\Theta) \to
\Re_+$ {can be chosen as} a distance on the space of measures supported
on $\Theta$. Since
the posterior is usually intractable, the expectation in
~\eqref{eq:posterior_expected_loss} is approximated using {posterior samples generated by Monte Carlo methods such as Markov chain Monte Carlo (MCMC) posterior simulation}. Given $M$ posterior Monte Carlo samples, 
$G_1, \ldots, G_M \sim p(G \mid Y_1, \ldots, Y_n)$, the optimization
problem can be approximated as 
\begin{align}
\label{eq:approximated_expected_loss}
    \Ghs = \text{arg}\min_{\Gh} \frac{1}{M} \sum_{m=1}^M \mathcal{D}(G_m, \Gh).
\end{align}
%
% After constructing the posterior similarity matrix of size $M \times M$, we can select the posterior sample with the minimum sum of distances to all other samples as an approximate solution to the optimization problem. Alternatively, a search algorithm equivalent to a barycenter problem can be employed.

\partitle{Algorithms.} We use a simple greedy procedure to solve the
optimization problem in
~\eqref{eq:approximated_expected_loss}. Specifically, we construct an
$M \times M$ matrix, where the entry in row $i$ and column $j$
represents the distance $\mathcal{D}(G_i, G_j)$ between the $i$-th and $j$-th posterior
samples. We then identify the index
$i^\star$ that minimizes the average distance across columns, 
$\frac{1}{M} \sum_{j=1}^M \mathcal{D}(G_i, G_j)$, and return
$G_{i^\star}$ as the greedy solution. This is equivalent to
selecting the best posterior sample from the posterior Monte Carlo
samples.  {We give a pseudo algorithm to describe the procedure in Algorithm~\ref{alg:summarization} in Appendix~\ref{subsec:appendix:algorithms}. }
% \mynote{need not be an MCMC sample - could be direct MC, or VB
%   or anything..}

\partitle{Truncation of mixing measures.} In practice, truncating the
mixing measure $G$ can accelerate computation, improve 
convergence and simplify implementation
\citep{ishwaran2001gibbs}.
 Specifically, we can use the
following truncated version of $G$: $\bar{G} = \sum_{k=1}^K \alpha_k
\delta_{\theta_k}$ with $0 < K < \infty$. {From Theorem 1 in \citet{ishwaran2001gibbs}, we can choose the truncation level based on the moments of the random weights. For example, under a Dirichlet process prior on $G$, we can choose $K$ such that $(\alpha_0/(\alpha_0+1))^{K-1}$ is small enough ($\alpha_0$ is the concentration parameter).} Ideally, we want to choose
the largest feasible value of $K$ to minimize the approximation error
introduced by truncation. % Once the measures are truncated,
We then
approximate the distance between two measures by the distance between
the truncated versions, i.e., $\mathcal{D}(G_1, G_2) \approx
\mathcal{D}(\bar{G}_1, \bar{G}_2)$. {Alternatively, one could proceed with random truncation as discussed in~\cite{griffin2016adaptive} or use a slice sampler~\citep{kalli2011slice}. Any alternative method could be used, as long as it provides a posterior Monte Carlo sample of the mixing measure .  Moreover, the proposed approach does not require matching truncation levels across Monte Carlo samples.  Under marginal MCMC~\citep{neal2000markov}, the algorithm can proceed with the empirical distribution of an imputed sample from $G$. Under a marginal algorithm, the latter can be generated. }

\partitle{Point estimate of the density function and partition.}
A point estimate of the mixing measure $\Gh$, implies
{a} point estimate of the density by convolution
with the kernel as
$
    \Fh = f * \Gh.
$
To estimate the partition, we determine the cluster membership
indicator $z_i$ by maximum a posteriori (MAP): 
\begin{align}
\label{eq:MAP_partition}
  \hat{z}_i = \argmax_{k \in \{1,\ldots,K\}} p(z_i = k \mid \Gh, y_i).
\end{align}
Here {$y_i$ generically denotes an observed data point}.
{Under the point estimate of the truncated mixing measure}, i.e., $\Gh \approx \sum_{k=1}^K \hat{w}_k
\delta_{\hat{\theta}_k}$, this becomes: 
\begin{align}
\label{eq:MAP_partition_truncation}
\hat{z}_i =  \argmax_{k \in \{1,\ldots,K\}} p(z_i = k \mid \hat{w}_1, \ldots, \hat{w}_K, \hat{\theta}_1, \ldots, \hat{\theta}_K, y_i).
\end{align}
While there are alternative methods to obtain a point estimate
$\hat{z}$  given the mixing measure {such as sampling or using a decision theoretic framework with Monte Carlo samples from $p(z_i \mid \hat{w}_1, \ldots, \hat{w}_K, \hat{\theta}_1, \ldots, \hat{\theta}_K, y_i)$}, we prefer the MAP as a
natural and computationally efficient approach.  {In addition, it is possible to extend the framework to summarize the density $F$ directly. However, such an approach would not be computational efficient since it might require evaluating the density on a grid of exponentially increasing size with increasing dimension. Moreover, obtaining a point estimate of the random partition would become complicated.}

% \partitle{Distances between measures as a loss function.}
 In the optimization in
\eqref{eq:approximated_expected_loss}, 
after truncation, {we require distances for pairs of} discrete mixing measures that may
have disjoint supports. Traditional
$f$-divergences~\citep{ali1966general}, such as the Kullback–Leibler
(KL) divergence, Jensen–Shannon (JS) divergence, and others, cannot directly be
used because they require access to the density ratio, which
may be undefined in this context. Consequently, optimal transport (OT)
metrics become a natural choice in this scenario. We will briefly discuss currently available OT
distances.

% \paragraph{Credible Ball.}  To characterize the uncertainty in the point estimate $\Gh$, we can define  a
%  credible ball of a given credible level $1-\alpha$, $\alpha \in [0,1]$: $B_{\epsilon^\star}(\Gh)$

\subsection{Optimal Transport Distances for the loss function}
\label{subsec:otdistances}

Let $G_1, G_2 \in \PP(\Theta)$ and $c: \Theta \times \Theta \to \Re^+$ be a ground metric. The Wasserstein-$p$ ($p \geq 1$) distance~\citep{villani2009optimal} between two measures $G_1$ and $G_2$ is defined as follows:
\begin{align}
\label{eq:Wasserstein}
    W_c^p(G_1, G_2) = \inf_{\pi \in \Pi(G_1, G_2)} \int_{\Theta \times \Theta} c(x, y)^p \, \mathrm{d}\pi(x, y),
\end{align}
where $\Pi(G_1, G_2) = \left\{\pi \in \PP(\Theta \times \Theta) \mid \pi(A, \Theta) = G_1(A), \ \pi(\Theta, B) = G_2(B) \ \forall A, B \subset \Theta \right\}$ is the set of all transportation plans/couplings. When $G_1$ and $G_2$ are discrete measures, i.e., $G_1 = \sum_{i=1}^{K_1} \alpha_i \delta_{x_i}$ and $G_2 = \sum_{j=1}^{K_2} \beta_j \delta_{y_j}$, the Wasserstein distance can be rewritten as:
\begin{align}
\label{eq:discrete_Wasserstein}
    W_c^p(G_1, G_2) = \min_{\pi \in \Gamma(\alpha, \beta)} \sum_{i=1}^{K_1} \sum_{j=1}^{K_2} c(x_i, y_j)^p \pi_{ij},
\end{align}
where the set of transportation plans becomes $\Gamma(\alpha, \beta) =
\left\{\pi \in \Re_+^{K_1 \times K_2} \mid \pi \mathbf{1} =
  \beta, \ \pi^\top \mathbf{1} = \alpha \right\}$. {In other words, Wasserstein distance is defined by selecting a coupling/copula $\pi$ that links the two marginal distributions. Each possible copula is scored by an implied cost of moving probability masses between the atoms/supports of the two discrete measures. Wasserstein distance is then the cost under the optimal, minimum cost copula.} The computation of
Wasserstein distance is often performed using linear
programming~\citep{peyre2020computational}, with a time complexity of
$\mathcal{O}((K_1 + K_2)^3 \log(K_1 + K_2))$. Alternatively, it can be
approximated using the entropic regularization
approach~\citep{cuturi2013sinkhorn}, which has a time complexity of
$\mathcal{O}\left(K_1 K_2 \log(K_1 + K_2) / \epsilon\right)$, where
$\epsilon > 0$ is the precision level. Consequently, using 
Wasserstein distance becomes impractical for large values of $K_1$ or
$K_2$.  {There is one important exception: Wasserstein distance is straightforward in closed-form for univariate measures. This motivates the next construction. When the ground metric $c=\|x-y\|_p$ ($p \geq 1$) is used, we denote $W_c^p(G_1, G_2)$ as $W_p^p(G_1, G_2)$.}
% Therefore, using the Wasserstein distance might
%  in an undesirable way 
% limit the truncation level $K_1$ and $K_2$.

%The Sliced Wasserstein
SW distance exploits the availability of a closed-form expression for
Wasserstein distance on the real line.
Specifically, for two distributions \( G_1, G_2 \in
\PP(\Re) \), the Wasserstein-\( p \) distance  is expressed as
\citep{peyre2020computational}: 
\begin{align}
\label{eq:1DW}
    W_p^p(G_1,G_2) = \int_{0}^1 |\CDF^{-1}_{G_1}(t) - \CDF^{-1}_{G_2}(t)|^p dt,
\end{align}
where \( \CDF^{-1}_{G_1} \) denotes the
inverse cumulative distribution functions (quantile functions) of \(
G_1 \). When $G_1=\sum_{i=1}^{K_1} \alpha_i \delta_{x_i}$ and  $G_2=\sum_{j=1}^{K_1}
\beta_j \delta_{y_j}$ are discrete 
 and assuming that the atoms of the two measures are sorted, the inverse CDFs  can
be evaluated as
\begin{align*}
  \CDF_{G_1}^{-1} (t) = \sum_{i=1}^{K_1} x_i
  I\left(\sum_{j=1}^{i-1}\alpha_j\leq t \leq \sum_{j=1}^{i}\alpha_j\right),
  \quad \CDF_{G_2}^{-1} (t) = \sum_{j=1}^{K_2} y_j I\left(\sum_{i=1}^{j-1}\beta_i\leq t \leq \sum_{i=1}^{j}\beta_i\right)
\end{align*}
Solving ~\eqref{eq:1DW} in this case has the time complexity of $\mathcal{O}((K_1+K_2)\log(K_1+K_2))$.

To leverage the closed-form solution of one-dimensional Wasserstein
distance in high-dimensional settings, SW distance
\citep{bonneel2015sliced} is introduced. The central idea of SW is to
randomly project two original measures onto  one-dimensional
measures and then compute the expected value of the one-dimensional
Wasserstein distance between the two projected measures. 
Conventional SW {($\Theta = \Re^d$)} employs a linear projection, defined as \( P_v(x) =
\langle v, x \rangle \) for \( v \in \mathbb{S}^{d-1} \), where \( v
\) represents the projection direction. The sliced Wasserstein-\( p \)
distance (\( p \geq 1 \)) between \( G_1 \) and \( G_2 \), using the
ground metric \( c(x,y) = |x-y| \), is defined as follows: 
\begin{align}
\label{eq:SW}
SW_p^p(G_1, G_2) = \mathbb{E}_{v \sim \mathcal{U}(\mathbb{S}^{d-1})}[W_p^p(P_v \sharp G_1, P_v \sharp G_2)],
\end{align}
where \( P_v \sharp G_1 \) and \( P_v \sharp G_2 \) are the push-forward measures of \( G_1 \) and \( G_2 \) through the function \( P_v \), and \( \mathcal{U}(\mathbb{S}^{d-1}) \) denotes the uniform distribution over the unit hypersphere.  {For the definition of push-forward measure, given two measurable spaces $(\Theta_1,\mathcal{X}_1)$ and
$(\Theta_2,\mathcal{X}_2)$, a measurable function $f:\Theta_1 \to
\Theta_2$, and a measure $G:\mathcal{X}_1 \to [0,\infty)$, the
pushforward of $G$ through $f$ is $f\sharp G(B) = G(f^{-1}(B))$
for any $B\in \mathcal{X}_2$. In the discrete setting with $G_1=\sum_{i=1}^{K_1} \alpha_i \delta_{\theta_i}$, $P_v\sharp G_1=\sum_{i=1}^{K_1} \alpha_i \delta_{P_v(\theta_i)}$ . Eq~\eqref{eq:SW} formalizes the expectation with respect to the projection $P_v$ of the univariate Wasserstein distances.}

The expectation in SW distance is intractable, thus Monte Carlo estimation is often employed to approximate SW. Specifically, let \( v_1, \ldots, v_L \overset{i.i.d.}{\sim} \mathcal{U}(\mathbb{S}^{d-1}) \) represent the projecting directions. The Monte Carlo estimation of SW is given by:
\begin{align}
\label{eq:MC_SW}
\widehat{SW}_p^p(G_1, G_2) = \frac{1}{L} \sum_{l=1}^L W_p^p(P_{v_l} \sharp G_1, P_{v_l} \sharp G_2).
\end{align}
% It is worth noting that variance reduction techniques can also be used~\citep{nguyen2023control,nguyen2024quasimonte,leluc2024slicedwasserstein}.
The overall time complexity  of SW evaluation is composed of the time required for
sampling projecting directions, the time for applying the projection
operator \( P_v \), and the time for computing one-dimensional
Wasserstein distances. When \( G_1 \) and \( G_2 \) are discrete
measures supported by \( K_1 \) and \( K_2 \) atoms, respectively, and \( L
\) Monte Carlo samples are used, the time complexity of SW is: 
$
\mathcal{O}(Ld + Ld(K_1 + K_2) + L(K_1 + K_2)\log(K_1 + K_2)) = \mathcal{O}(Ld(K_1 + K_2) + L(K_1 + K_2)\log(K_1 + K_2)),
$
where \( \mathcal{O}(Ld) \) accounts for sampling projecting
directions, \( \mathcal{O}(Ld(K_1 + K_2)) \) is for the projection,
and \( \mathcal{O}(L(K_1 + K_2)\log(K_1 + K_2)) \) is for computing \(
L \) one-dimensional Wasserstein distances. Additionally, the
projection complexity of SW is \( \mathcal{O}(Ld) \), which
corresponds to the memory required for storing the \( L \) projecting
directions. In summary, SW is very scalable in terms of  $K_1$ and $K_2$, having {i.e., it has near linear complexity in terms of $K_1$ and $K_2$} . It allows accurate
truncation with large $K_1$ and $K_2$.

{A potentially interesting alternative to SW is the multivariate Cramer distance~\citep{baringhaus2004new,baringhaus2010rigid}. Despite being more scalable than  Wasserstein distance, it is not as geometrically meaningful since it does not involve finding the optimal coupling between two measures. Moreover, by using entropic regularization, the Wasserstein distance can be seen as a generalization of  Cramer distance by forcing the coupling to be an independent coupling~\citep{feydy2019interpolating}. In addition,  Cramer distance is not as scalable as SW. Cramer distance has quadratic complexity with respect to $K_1$ and $K_2$, while SW has near-linear complexity. Furthermore, SW is also statistically scalable in dimension. In particular, from~\citet{nadjahi2020statistical,nguyen2021distributional,boedihardjo2025sharp}, we have that
$
\mathbb{E}\left[\left|SW_p(\Gh_{1,K_1}, \Gh_{2,K_2}) - SW_p(G_1, G_2)\right|\right] = \mathcal{O}(K_1^{-1/2} + K_2^{-1/2})
$
(no exponential relationship between sample size and dimension) for $G_1, G_2 \in \mathcal{P}_p(\mathbb{R}^d)$, where $\Gh_{1,K_1}, \Gh_{2,K_2}$ are empirical distributions over $K_1$ and $K_2$ i.i.d. samples from $G_1$ and $G_2$, respectively. Overall, SW is an attractive option in terms of computational benefits.
 }

\section{Sliced Optimal Transport Distances for Gaussian Mixing Measures}
\label{sec:SWGMM}

{For the mixing measures \( G_1 \) and \( G_2 \) in mixtures of Gaussians, }
we use parameter space \( \Theta = \Re^d \times
S_d^{++}(\Re) \), where \( S_d^{++}(\Re) \) is the
manifold of all symmetric positive definite matrices.
Conventional SW distance can not directly be applied in this
context, as it is defined for measures on vector spaces.
% In this section,
We first discuss an approach to apply SW using vectorization in
Section~\ref{subsec:vSW}.
We then propose two novel variants of SW that preserve geometry.
In Section~\ref{subsec:mixedSW} we start with a new variant of SW based
on generalized geodesic projection onto the product of manifolds, which we call
Mixed SW (Mix-SW).
Finally, we introduce another variant of SW for finite Gaussian mixing
measures by comparing their induced mixture measures, which we call sliced mixture Wasserstein (SMix-W). {Finally, we note again that in case of a focus on the implied density only, restriction to (less parsimonious, but still flexible) location mixtures is a viable alternative.}

% As discussed in the previous section, Wasserstein distance is computationally expensive in terms of the number of supports for the mixing measures. Therefore, we cannot use the Wasserstein distance when we want to approximate effectively the mixing measure via truncation.  In this section, we discuss the usage of slicing approaches to obtain more computationally scalable distances. 

\subsection{Vectorized Sliced Wasserstein}
\label{subsec:vSW}
We are aiming to compare measures on the product
of the Euclidean manifold and the manifold of symmetric positive
definite (SPD) matrices, denoted as \( \Theta = \Re^d \times
S_d^{++}(\Re) \), using the SW
distance. However, SW is defined on vector
spaces. A straightforward approach is to convert measures
on \( \Re^d \times S_d^{++}(\Re) \) into measures on a
vector space.  For any \( \theta = (\mu, \Sig) \in
\Re^d \times S_d^{++}(\Re) \), we can arrange the
entries of \( \Sig \) to obtain a vector representation, which can
then be stacked with \( \mu \). For simplicity, we define the
transformation \( V(\theta) = (\mu, \Sig^{(1)}, \ldots,
\Sig^{(d)}) \), where \( \Sig^{(i)} \) is the \( i \)-th row of
the matrix \( \Sig \).  With this transformation, we can redefine
SW distance as
\begin{align}
    \label{eq:vecSW}
    SW_p^p(G_1,G_2) = \mathbb{E}_{v \sim \mathcal{U}(\mathbb{S}^{d(d+1)-1})}[W_p^p (P_v \sharp V \sharp G_1, P_v \sharp V \sharp G_2)],
\end{align}
for any \( G_1, G_2 \in \PP(\Re^d \times S_d^{++}(\Re)) \). 
Despite the attractive simplicity of this approach, there are two main
complications to
consider. The first is that vectorization destroys the geometry of the
space, which may result in a distance that lacks geometric meaning. In
contrast to the Wasserstein distance, where the ground metric can be
flexibly designed, the ground metric in SW is constrained to exist
in a one-dimensional space. The second issue pertains to the
high-dimensional projection direction space, \( \mathbb{S}^{d(d+1)-1}
\), which may require increased computation and memory to achieve
accurate approximations via Monte Carlo estimation.  The time
complexity of vectorized SW is given by  $ 
\mathcal{O}(L d^2 (K_1 + K_2) + L (K_1 + K_2) \log (K_1 + K_2))$
and the projection complexity is $
\mathcal{O}(L d^2)$,
both of which are quadratic in dimensions. {Alternatively, we can also utilize the transformation $V(\theta)=(\mu,B)$ with $B$ as the vectorized version of the upper or lower half of $\Sigma$ or any triangular factorization. However, the complexity still remains at least quadratic in terms of dimensions. It is worth noting that the vectorization transformation is generic in the sense that it can work with other spaces without a symmetric structure. We present a pseudo algorithm for the vectorization approach in Algorithm~\ref{alg:SW} in Appendix~\ref{subsec:appendix:algorithms}.}

\subsection{Mixed Sliced Wasserstein}
\label{subsec:mixedSW}
To address the loss of geometric information, we propose a new
variant of SW distance using geodesic
projection of the product manifolds \( \Re^d \times S_d^{++}(\Re)
\). In brief, we define a notion of projecting a support point $(\mu,\Sigma) \in  \Re^d \times S_d^{++}(\Re)
$ onto a curve with  associated velocity vector $V_w =(w_1v,w_2A)$ {($v$ and $A$ are unit vectors in an appropriate sense which will be defined later)} in a way that the projection is easy to evaluate. Using the projection, the desired SW is defined as the expectation of the projected one-dimensional Wasserstein distance, taking the expectation with under the uniform-law over the random curve parameters $(v,A,w)$.
% distance. 
% \mynote{K: would it be helpful to provide here a brief ``road map''?
%   I tried - but not sure at all :-)}
  
% In summary, we define a notion of projecting $(\mu,\Sig)$ onto a
% curve indexed by $(v,A,w)$, in a way that (i) the projection is easy
% to evaluate; and (ii) averaging over $(v,A,w)$ defines the desired SW
% distance. 
We begin by reviewing some basic definitions relevant to
Riemannian manifolds, including the inner product, geodesics, length,
geodesic distance, and the exponential map, as detailed in
Appendix~\ref{subsec:manifold_review}. And, we review
the concept of geodesic projection~\citep{bonet2024sliced} and explore
certain properties of the manifold of symmetric positive definite
matrices \( S_d^{++}(\Re) \)~\citep{pennec2019riemannian} in Appendix~\ref{subsec:appendix:mixSW}.

% which plays a vital role in the development of our SW variant.

The geodesic projection
and the corresponding SW distance for measures on \( S_d^{++}(\Re) \)
have been investigated in~\citet{bonet2023sliced}. While the geodesic
projection for the product of manifolds was introduced
in~\citet{bonet2024sliced}, it has not been explicitly derived for any
specific case. In this work, we extend the notion of geodesic
projection to a generalized geodesic projection by projecting onto a
generalized curve with adjusted velocity vectors for each marginal
manifold. This adjustment is essential for achieving the identity of
indiscernibles {(see Appendix~\ref{subsec:proof:theorem:mixSW_metric})} in the subsequently defined SW metric.

\begin{figure}[!t]
\begin{center}
  \begin{tabular}{c} 
\widgraph{0.6\textwidth}{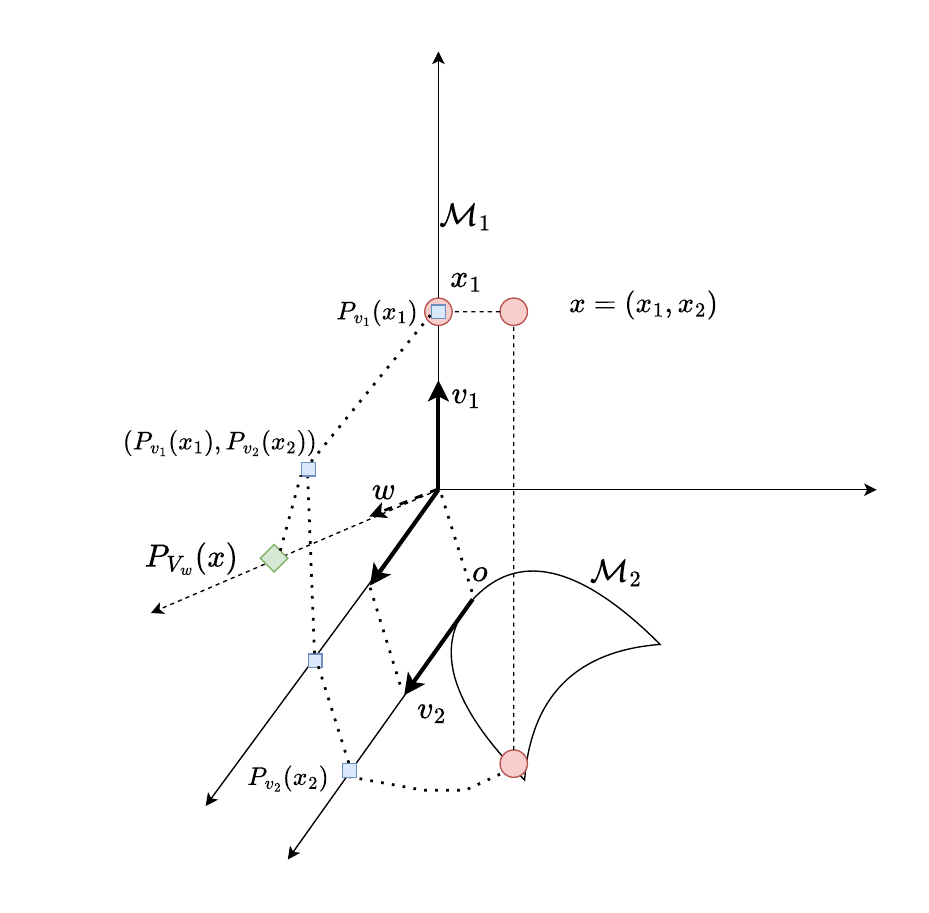} 

  \end{tabular}
  \end{center}
  \vspace{-0.3 in}
  \caption{
  \footnotesize{ {The figure illustrates the computation of generalized geodesic projection of $x=(x_1,x_2)$ (red ball) in  $\mathcal{M}_1\times \mathcal{M}_2$ ($\mathcal{M}_1$ is  a line (1D Eculidean), $\mathcal{M}_2$ is embedded in 2D with curvature). Blue squares mark the marginal projections for $x_1$ and $x_2$.  The final generalized geodesic projection is the green diamond. The green diamond represents the length from the origin to the projected point on the product manifold (the projected point itself is not shown in the figure).}
}
} 
  \label{fig:projection}
\end{figure}

\begin{definition}
    \label{def:generalized_geodesic_projection}
    Given a product manifolds $\mathcal{M}_1\times \mathcal{M}_2$ with the origin $o= (o_1,o_2)$,  a generalized curve passing through the origin with the velocity vector $V_w =(w_1v_1,w_2v_2)$ with $\langle v_1,v_1\rangle_{o_1} = 1$, $\langle v_2,v_2\rangle_{o_2}=1$, and $w_1^2+w_2^2=1$, the generalized geodesic projection onto the generalized curve created by $V_w$ is defined as:
    \begin{align}
    \label{eq:geodesic_projection}
    P_{V_w}(x)= \argmin_{t \in \Re }c(x,\exp_o(tV_w)),
\end{align}
where $c$ is {a} geodesic distance  on the product manifold.
\end{definition}
{We
illustrate the intuition of the generalized geodesic projection in
Figure~\ref{fig:projection}. Note that the generalized geodesic projection in \eqref{eq:geodesic_projection} is defined as  the length $(\argmin_t)$ of the curve from the origin to the closest point on the curse from $x$.} The generalized geodesic projection has a closed-form
expression on \( \Re^d \times S_d^{++}(\Re) \) as in the following proposition:
\begin{proposition}
    \label{prop:generalized_geodesic_projection} The generalized
    geodesic projection  onto a generalized curve passing through the
    origin with the velocity vector $V_w=(w_1 v,w_2A)$
    ($\|v\|_2^2=1,A \in S_d(\mathbb{R}),\|A\|_F^2=1, w_1^2+w_2^2=1$) on the product manifold
    of $\Re^d \times S_d^{++}(\Re)$ has the following form: 
    $$
        P_{V_w}((\mu,\Sig)) =  w_1 \langle \mu,v\rangle+ w_2Trace(A\log \Sig).
    $$
\end{proposition}
The proof of Proposition 1 is given in
Appendix~\ref{subsec:proof:prop:generalized_geodesic_projection}. The generalized geodesic projection {(the green diamond in Figure~\ref{fig:projection})} is a weighted
combination of the geodesic projection on the Euclidean manifold and
the geodesic  projection  on the  manifold $S_d^{++}(\Re)$. Using the generalized geodesic projection, we now define the mixed Sliced Wasserstein (Mix-SW) distance.

\begin{definition}
\label{def:mixSW}
    Given two measures $G_1$ and $G_2$ in $\PP(\Re^d
    \times S_d^{++}(\Re))$, $p\geq 1$,  the Mixed Sliced Wasserstein
    distance is defined  as follows: 
\begin{align}
\label{eq:mixsw}
    \text{Mix-}SW_p^p(G_1,G_2) =\mathbb{E}_{(w,v,A)\sim \mathcal{U}(\mathbb{S}) \otimes \mathcal{U}(\mathbb{S}^{d-1}) \otimes \mathcal{U}(S_d(\Re)) } [W_p^p(P_{V_w} \sharp G_1, P_{V_w}\sharp G_2)],
\end{align}
where $V_w=(w_1v,w_2A)$ and $\mathcal{U}(\mathcal{X})$ is the uniform distribution over the set $\mathcal{X}$.
\end{definition}
Recall that $P_{V_w}$ was defined as a length (on the curve) in \eqref{eq:geodesic_projection} implying $W_p^p$ in \eqref{eq:mixsw} as distance. Mix-SW is similar to hierarchical hybrid sliced Wasserstein
(H2SW)~\citep{nguyen2024h2sw}. Both definitions combine projections from
marginals. However, Mix-SW comes from generalized geodesic projection
for a specific product of manifold while H2SW arises from randomly
combining two general types of Radon transforms.  Also, H2SW is
only introduced for the product of the Euclidean manifold and the
hypersphere. {We first show that $\text{Mix-}SW_p^p(G_1,G_2)$ is bounded as long as
the expected geodesic distances with respect to $G_1$ and $G_2$ to any
point $(\mu_0,\Sig_0)$ are bounded. }

\begin{proposition}
    \label{prop:boundness}  If $\int_{\Re^d \times S_d^{++}(\Re)}c((\mu_1,\Sig_1),(\mu_0,\Sig_0))^p  \mathrm{d}G_1(\mu_1,\Sig_1) < \infty$ \\ and $\int_{\Re^d \times S_d^{++}(\Re)}c((\mu_0,\Sig_0),(\mu_2,\Sig_2))^p  \mathrm{d}G_2(\mu_2,\Sig_2) <\infty$ for any $(\mu_0,\Sig_0)\in\Re^d \times S_d^{++}(\Re)$ with $ c((\mu_1,\Sig_1),(\mu_2,\Sig_2) = \sqrt{\|\mu_1-\mu_2\|_2^2+ \|\log \Sig_1 -\log \Sig_2\|_F^2},$ then $\text{Mix-}SW_p^p(G_1,G_2)< \infty$.
\end{proposition}
The proof of Proposition~\ref{prop:boundness} is given in
Appendix~\ref{subsec:proof:prop:boundness}. Next, we  show that
Mix-SW is a valid metric.

\begin{theorem}
    \label{theorem:mixSW_metric} Mixed Sliced Wasserstein is a valid metric on the space of measures which belong to $\PP(\Re^d \times S_d^{++}(\Re))$ and satisfy the constraint in Proposition~\ref{prop:boundness}.
\end{theorem}

The proof of Theorem~\ref{theorem:mixSW_metric} is given in
Appendix~\ref{subsec:proof:theorem:mixSW_metric} which extends the
technique of the proofs
in~\citet{bonnotte2013unidimensional},\citet{nadjahi2020statistical}, and \citet{bonet2023sliced}
with the usage of the generalized geodesic projection.

\begin{corollary}
    \label{corollary:mixed_gaussian} Mix-SW is also a metric on the space of finite mixture of Gaussians.
\end{corollary}

Corollary~\ref{corollary:mixed_gaussian} {is based on the identifiability of finite mixture of Gaussians (Proposition 2 in~\cite{yakowitz1968identifiability})} and it suggests that Mix-SW can 
also be used to compare two finite mixtures of Gaussians based on their
mixing measures on $\Re^d \times S_d^{++}(\Re)$.

On the computational side, the expectation in
Definition~\ref{def:mixSW} is intractable. 
We therefore employ Monte Carlo estimation to approximate 
Mix-SW. Specifically, we sample \( (w_1, v_1, A_1), \ldots, (w_L, v_L,
A_L) \overset{\text{i.i.d}}{\sim} \mathcal{U}(\mathbb{S}) \otimes
\mathcal{U}(\mathbb{S}^{d-1}) \otimes \mathcal{U}(S_d(\Re)) \). The
Monte Carlo estimate of Mix-SW is then defined as: 
\begin{align}
    \widehat{\text{Mix-}SW}_p^p (G_1, G_2) = \frac{1}{L} \sum_{l=1}^L W_p^p(P_{V_{w,l}} \sharp G_1, P_{V_{w,l}} \sharp G_2),
\end{align}
where \( V_{w,l} = (w_{l1} v_l, w_{l2} A_l) \). When \( G_1 \) and \( G_2 \) are discrete measures with \( K_1 \) and \( K_2 \) supports, respectively, the time complexity of Mix-SW is $
\mathcal{O}((K_1 + K_2 + L)d^3 + L(K_1 + K_2)d^2 + L(K_1 + K_2) \log (K_1 + K_2)),$
which arises from sampling \( A \sim \mathcal{U}(S_d(\Re)) \) (see Algorithm 1 in~\citet{bonet2023sliced}), computing the matrix logarithm, projecting the samples, and solving one-dimensional Wasserstein distances. The projection complexity of Mix-SW is \( \mathcal{O}(Ld^2) \) since it requires storing \( L \) projection matrices \( A_1, \ldots, A_L \). {We present a pseudo algorithm for computing Mix-SW in Algorithm~\ref{alg:MixSW} in Appendix~\ref{subsec:appendix:algorithms}. }

\subsection{Sliced Mixture Wasserstein}
\label{subsec:SMixW}
Mix-SW compares measures in $\PP(\Re^d \times
S_d^{++}(\Re))$. But it is not specifically designed for Gaussian
mixing measures. To leverage the structure of Gaussian mixing
measures, we introduce a variant called the sliced mixture Wasserstein
(SMix-W) distance, which compares Gaussian mixing measures via their
induced mixture of Gaussian measures. SMix-W is inspired by the
Mixture Wasserstein distance~\citep{delon2020wasserstein}, which is
defined specifically for comparing mixtures of Gaussian measures.  {We refer the reader to Appendix~\ref{subsec:appendix:smix} for a more detail discussion of the connection to Mixture Wasserstein. We  start with the linearity of mixture of Gaussians.} When  $G_1=\sum_{i=1}^{K_1} \alpha_i \delta_{(\mu_{1i},\Sig_{1i})}$ and $G_2=\sum_{j=1}^{K_2} \beta_j \delta_{(\mu_{2j},\Sig_{2j})}$, we have $F_1 = f * G_1:=\sum_{i=1}^{K_1} \alpha_i \mathcal{N}(\mu_{1i},\Sig_{1i})$ and $F_2  = f* G_2:=\sum_{j=1}^{K_2} \beta_j \mathcal{N}(\mu_{2j},\Sig_{2j})$. For a vector $v \in \Re^d$ and $P_v(x) = \langle x,v\rangle$, we have $P_v\sharp F_1 :=\sum_{i=1}^{K_1} \alpha_i \mathcal{N}(\langle v,\mu_{1i}\rangle,v^\top \Sig_{1i}v) $ and $P_v\sharp F_2 :=\sum_{j=1}^{K_2} \beta_j \mathcal{N}(\langle v,\mu_{2j}\rangle,v^\top \Sig_{2j}v)$, which are two one-dimensional Gaussian mixtures with the mixing measures $P'_v \sharp G_1$ and $P'_v \sharp G_2$ with $P'_v(\mu,\Sig) = (\langle v,\mu\rangle, v^\top \Sig v)$. After applying linear
projection to the mixture Gaussians (or $P'_v$ on the Gaussian mixing
measure), we can use generalized geodesic projections to obtain the sliced Mixture Wasserstein as follows:
% $P_{v,w}(\mu,\Sig) = w_1 \langle v,\mu \rangle+ w_2 \log(\sqrt{v^\top\Sig v})$.

\begin{definition}
\label{def:smixW}
    Given two finite discrete measures $G_1$ and $G_2$ in
$\PP(\Re^d \times S_d^{++}(\Re))$, $p\geq 1$, and $P_{v,w}(\mu,\Sig) =
w_1 \langle v,\mu \rangle+ w_2 \log(\sqrt{v^\top\Sig v})$, the sliced
mixture Wasserstein (SMix-W) is defined as follows:
\begin{align*}
    \text{SMix-}W_p^p(G_1,G_2) =\mathbb{E}_{(w,v)\sim \mathcal{U}(\mathbb{S}) \otimes \mathcal{U}(\mathbb{S}^{d-1}) } [W_p^p(P_{v,w} \sharp G_1, P_{v,w}\sharp G_2)],
\end{align*}
where $\mathcal{U}(\mathcal{X})$ is the uniform distribution over the set $\mathcal{X}$.
\end{definition}
 Compared to SW and Mix-SW, the projection space of SMix-W is smaller, specifically \(\mathbb{S}^{d-1} \times \mathbb{S}\), as it utilizes a single projecting direction \(v\) for both \(\mu\) and  \(\Sig\).

% \begin{proposition}
%     \label{prop:smixw_properties} 
%     (i) 
% \end{proposition}

\begin{proposition}
    \label{prop:boundness_smix} If $\int_{\Re^d \times S_d^{++}(\Re)}c((\mu_1,\Sig_1),(\mu_0,\Sig_0))^p  \mathrm{d}G_1(\mu_1,\Sig_1) < \infty$ and \\ $\int_{\Re^d \times S_d^{++}(\Re)}c((\mu_0,\Sig_0),(\mu_2,\Sig_2))^p  \mathrm{d}G_2(\mu_2,\Sig_2) <\infty$ for any $(\mu_0,\Sig_0)\in\Re^d \times S_d^{++}(\Re))$ with $ c((\mu_1,\Sig_1),(\mu_2,\Sig_2)) = \sqrt{\|\mu_1-\mu_2\|_2^2+ 0.25\log(\lambda_{max}(\Sig_1,\Sig_2))^2}$ ($\lambda_{max}(\Sig_1,\Sig_2)$ is the  largest eigenvalue of the generalized problem $\Sig_1 v = \lambda \Sig_2 v$), then $\text{SMix-}W_p^p(G_1,G_2)< \infty$.
\end{proposition}

The proof of Proposition~\ref{prop:boundness_smix} is given in Appendix~\ref{subsec:proof:prop:boundness_smix}. After showing the {SMix-W} is bounded, we show that SMix-W is a valid metric for discrete measures on $\PP(\Re^d \times S_d^{++}(\Re))$.

\begin{theorem}
    \label{theorem:metricity_smixw} SMix-W is a valid metric on the space of finite discrete measures on $\PP(\Re^d \times S_d^{++}(\Re))$ which satisfy the constraint in Proposition~\ref{prop:boundness_smix}.
\end{theorem}
The proof of Theorem~\ref{theorem:metricity_smixw} is given in
Appendix~\ref{subsec:proof:theorem:metricity_smixw}. {The claim is established only for} finite discrete measures on
$\PP(\Re^d \times S_d^{++}(\Re))$  since the proof of identity of
indiscernibles of SMix-W relies on the identifiability of finite
mixture of Gaussians (Proposition 2
in~\cite{yakowitz1968identifiability}). In addition, from the
identifiability, SMix-W is also a metric between finite mixtures of
Gaussians measures. 

On the computational side, the expectation in Definition~\ref{def:smixW} is also intractable. As before, we use Monte Carlo estimation to approximate the value of SMix-W. In particular, we sample \((w_1, v_1), \ldots, (w_L, v_L) \overset{i.i.d.}{\sim} \mathcal{U}(\mathbb{S}) \otimes \mathcal{U}(\mathbb{S}^{d-1})\). The Monte Carlo estimation of SMix-W is defined as follows:
\begin{align}
    \widehat{\text{SMix-}W}_p^p (G_1,G_2) =  \frac{1}{L} \sum_{l=1}^L W_p^p(P_{v_l,w_l} \sharp G_1, P_{v_l,w_l} \sharp G_2).
\end{align}
When \(G_1\) and \(G_2\) are discrete measures with \(K_1\) and \(K_2\) supports, respectively, the time complexity of SMix-W is \(\mathcal{O}(L(K_1 + K_2 + 2)d^2 + L(K_1 + K_2) \log (K_1 + K_2))\), due to the computation of the projections and the solving of one-dimensional Wasserstein distances. The projection complexity of SMix-W is \(\mathcal{O}(Ld)\), as it only requires storing \(L\) projections of \((w_1, v_1), \ldots, (w_L, v_L)\).  We observe that SMix-W reduces the time complexity from \(\mathcal{O}(d^3)\) of Mix-SW to \(\mathcal{O}(d^2)\) by exploiting the linearity of mixtures of Gaussians. Compared to vectorized SW and Mix-SW with \(\mathcal{O}(d^2)\) in projection complexity, SMix-W has a better projection complexity of \(\mathcal{O}(d)\), making it more scalable with respect to the number of dimensions. Furthermore, a lower-dimensional projection space for SMix-W may lead to a reduced number of projections \(L\) required for a good approximation. {We present a pseudo algorithm for computing SMix-W in Algorithm~\ref{alg:SMixW} in Appendix~\ref{subsec:appendix:algorithms}.}

\section{Empirical Analysis}
\label{sec:experiments}
We  assess clustering and density estimation under 
two alternative  approaches {to summarize the posterior Monte Carlo samples}: summarizing random
partitions versus the proposed novel method of summarizing random mixing
measures. For the first approach, we utilize the SALSO
package~\citep{dahl2022search} with its greedy search algorithm to
obtain point estimates of the random partition using Binder, VI
, and omARI. {Below we will refer to these summaries as the ``alternative approaches"}. For the proposed new methods, we employ vectorized
SW, Mix-SW, and SMix-W (all are approximated with $L=100$ projections)
to obtain point estimates of the random mixing measures.  For
evaluation, we use Binder loss, VI loss, and omARI loss to assess
clustering performance, while we employ approximated Total Variation
(TV) and approximated SW (with $L=1000$ projections) computed on a
grid over the data space to evaluate the density estimates.  {Note that this setup favor approaches that use the same loss to summarize posterior Monte Carlo samples}.

The proposed approach  remains valid
for any  BNP mixture models with  arbitrary prior
and sampling models. For easier exposition and to facilitate the
comparison  
we work with the conjugate truncated Dirichlet process mixture of Gaussians models~\citep{ishwaran2001gibbs}:
% to have a simple inference process: 
\begin{align}
\label{eq:model}
    &\beta_1,\ldots,\beta_K\mid \alpha  \sim Beta(1,\alpha), \quad \quad 
    w_k = \beta_k \prod_{j=1}^{k-1} (1-\beta_j),\\
    &z_i|w_1,\ldots,w_K \sim Multinomial(w_1,\ldots,w_K),\nonumber \quad
    (\mu_i,\Sig_i) \mid \mu_0,\lambda, \Psi, \nu  \sim \mathcal{NIW}(\mu_0,\lambda, \Psi, \nu), \nonumber \\
   & y_{i} \mid  \mu_{1:K}, \Sig_{1:K},z_i=k  \sim \mathcal{N}(\mu_k,\Sig_k),
    ~~ i=1,\ldots,n, \nonumber 
  \end{align}
where $K>0$ is the truncation level, and $\mathcal{NIW}$ denotes the
Normal Inverse Wishart distribution. Inference under ~\eqref{eq:model} of the above model
can be carried out efficiently by a blocked Gibbs
sampler~\citep{ishwaran2001gibbs} to simulate from the joint distribution
$p(\beta_{1:K},\mu_{1:K},\Sig_{1:K},z_{1:n}\mid y_{1:n})$. We implement inference in {Python} for a  
% We run this model on
simulated dataset
in Section~\ref{subsec:simluated_data}, and for the Old Faithful geyser
dataset~\citep{azzalini1990look}  in Section~\ref{subsec:old_faithful}.  {In Appendix~\ref{subsec:MC_Approximation}, we report a brief sensitivity analysis for the Monte Carlo approximation of SW, Mix-SW, and SMix-W.}
% The blocked
% Gibbs sampler
%  defines transition probabilities defined by sampling from the
% following complete conditional posterior distributions 
% (1) $p(z_i=k \mid \beta_{1:K},\mu_{1:K},\Sig_{1:K},y_i) \propto w_k
% \mathcal{N}(y_i|\mu_k,\Sig_k)$;
% (2) $\beta_k \sim Beta(1+n_k,\alpha + \sum_{j>k} n_j)$; and 
% (3) $\mu_k,\Sig_k \sim \mathcal{NIW}\left(\frac{\lambda \mu_0+
%         n_k \bar{y}_k}{\lambda +n_k},\lambda+
%       n_k,\Psi+\sum_{i|z_i=k}(y_i-\bar{y}_k)(y_i - \bar{y}_k)^\top
%     \right. \\ \left. +\frac{\lambda n_k}{\lambda+n_k}(\bar{y}_k -
%       \mu_0)(\bar{y}_k - \mu_0)^\top ,\nu+n_k\right)$, 

% where $n_k$ is the number of members of the $k$-th clusters given a
% partition, $\bar{y}_k=\frac{1}{n_k}\sum_{i=1}^n y_{i}\delta_{z_i=k}$
% is the mean of cluster $k$-th.

\subsection{Simulated data}
\label{subsec:simluated_data}

 Let $V=1.5^2\cdot I_2$. 
We sample 200 i.i.d data 
$$
y_i \iid
\frac{1}{4}\NN\left((-2,-2),V)\right)   +
\frac{1}{4}\NN\left((2,-2),V)\right)+\frac{1}{4}\NN\left((-2,2),V\right)+\frac{1}{4}\NN\left((2,2),V)\right),
$$
$i=1,\ldots,n=200$. 
We run 10000 blocked Gibbs sampler iterations (9000
burn-in iterations) with the following hyperparameters
$\mu_0=(0,0),\Psi=diag((1,1)),\lambda=1,\nu=4,\alpha=1,K=100$. {We repeat the simulation 25 times.}
% \mynote{K: in simulation no need for burn-in :-) Convergence of an
%   ergodic Markov chain doesn't depend on the initial value.}

\begin{figure}[!t]
\begin{center}
    \begin{tabular}{cc}
  \widgraph{0.25\textwidth}{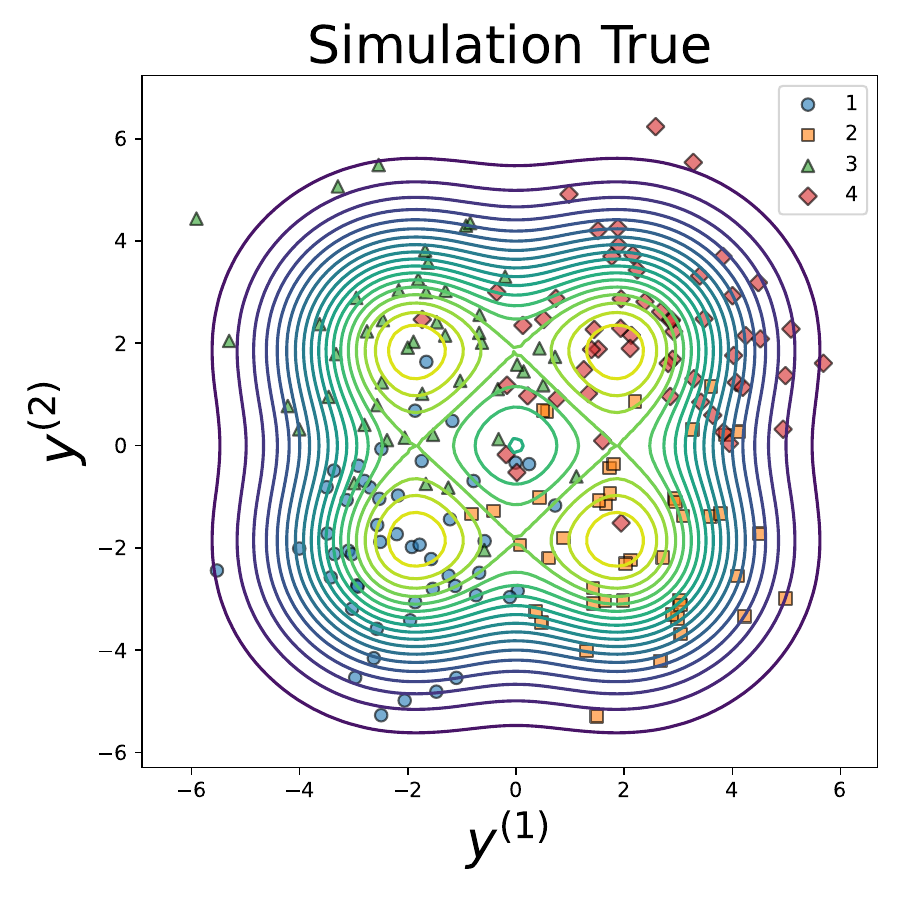} &\widgraph{0.25\textwidth}{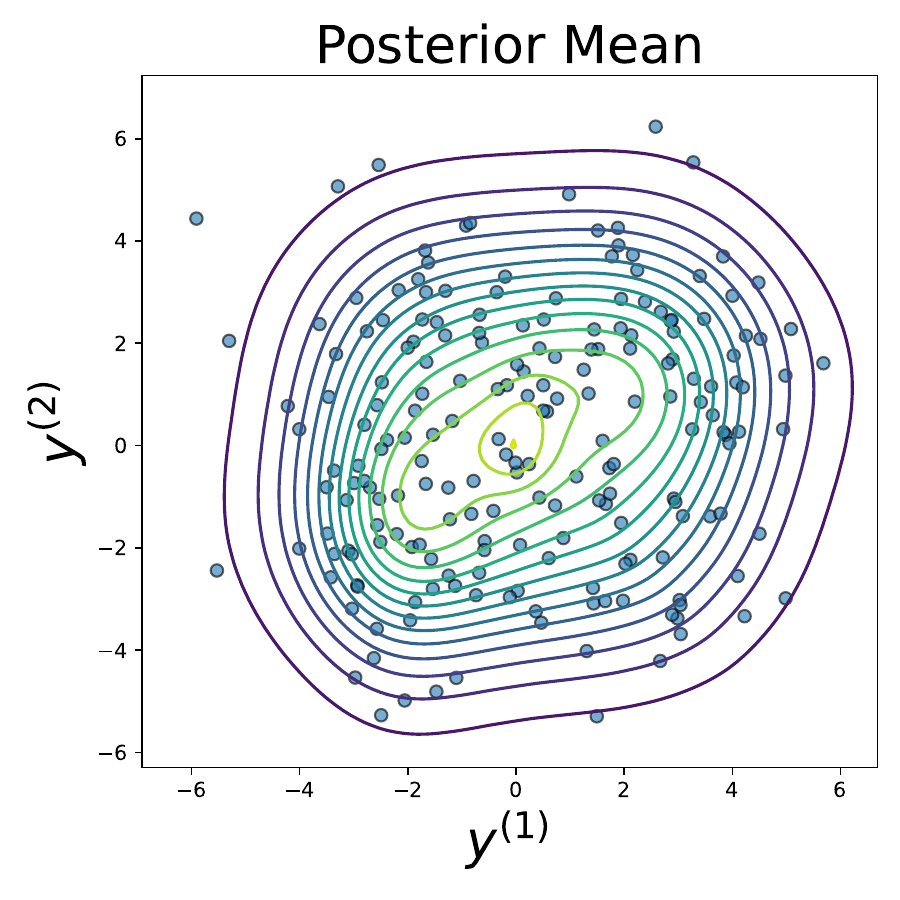}  

  \end{tabular}
  \begin{tabular}{ccc} 
\widgraph{0.25\textwidth}{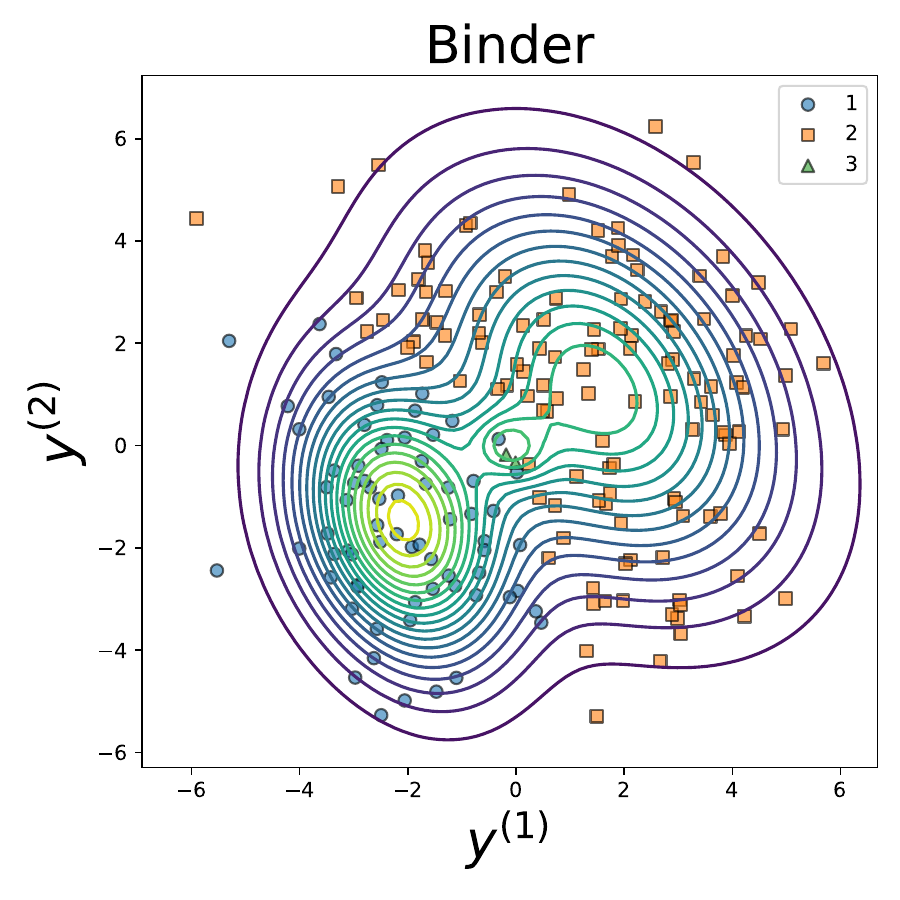} 
&\widgraph{0.25\textwidth}{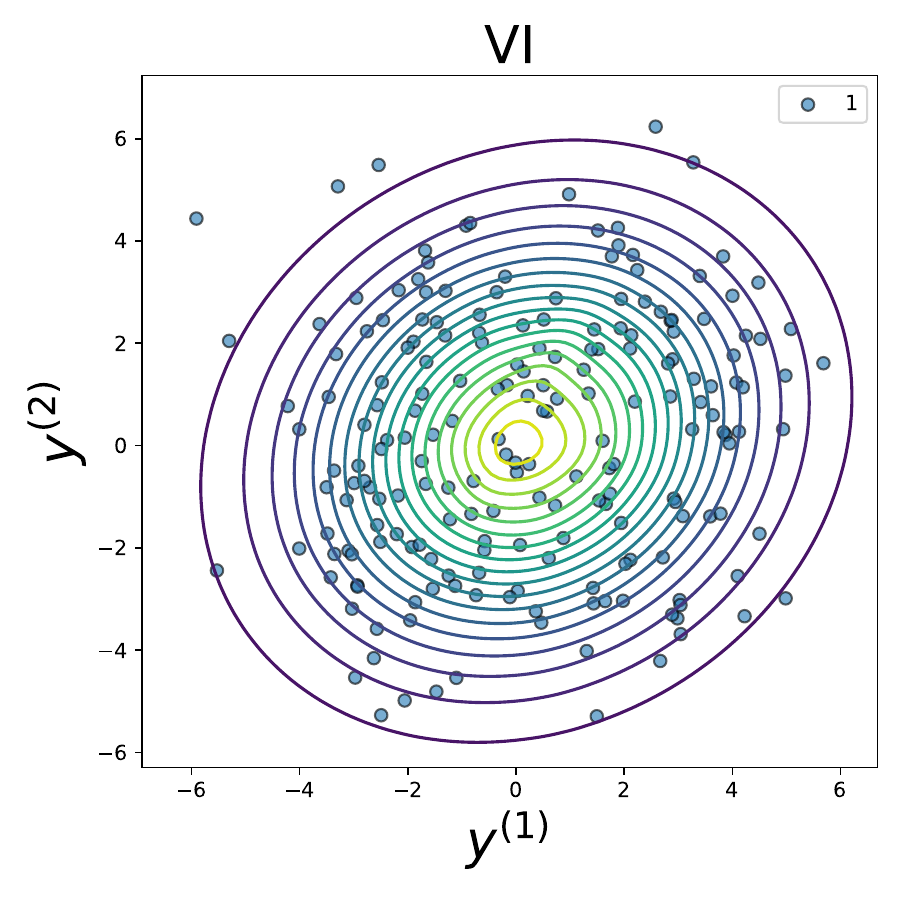} 
 &\widgraph{0.25\textwidth}{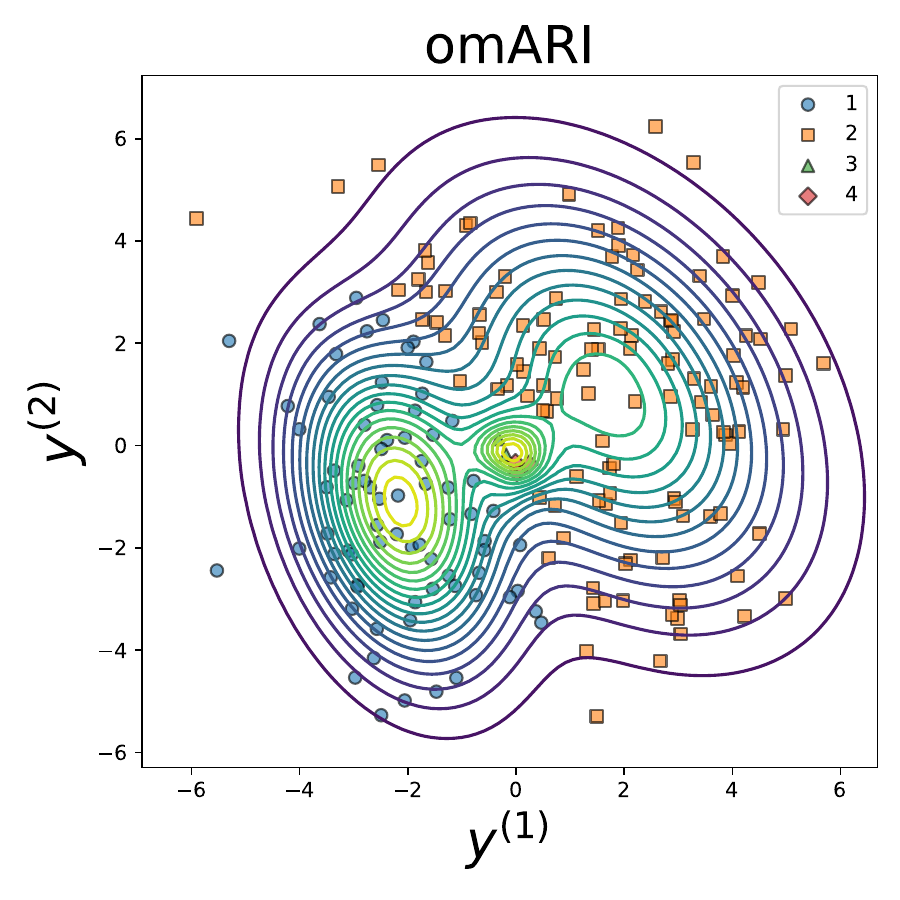} \\
 \widgraph{0.25\textwidth}{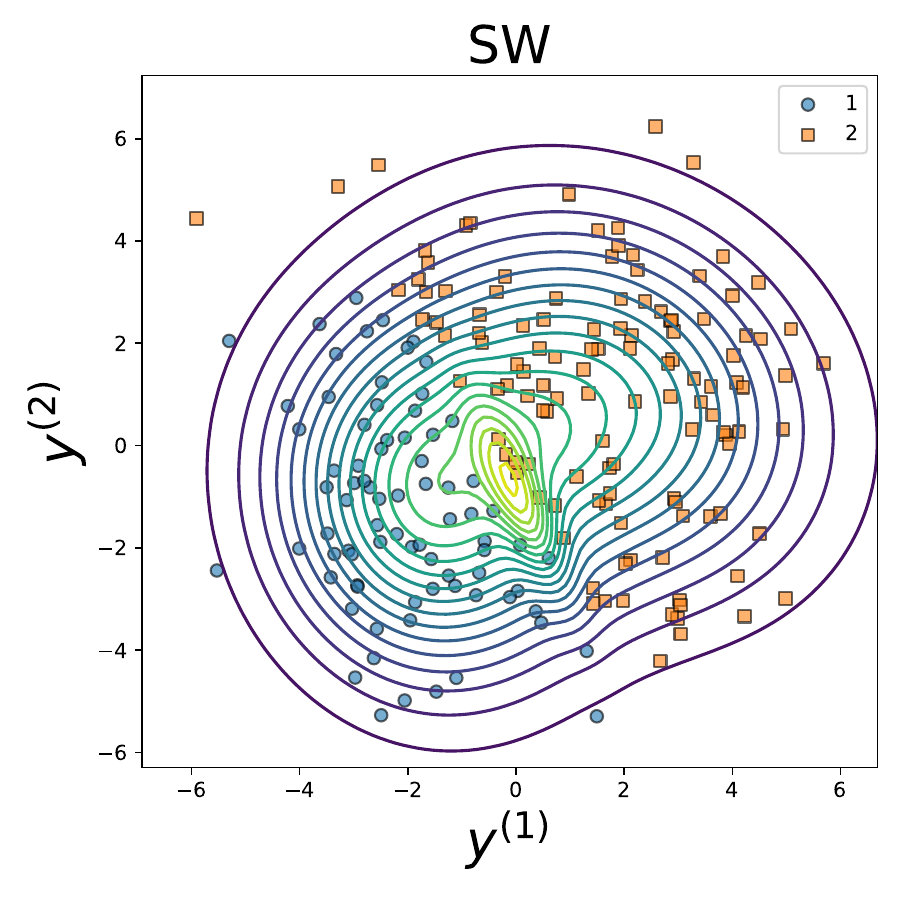}  
&\widgraph{0.25\textwidth}{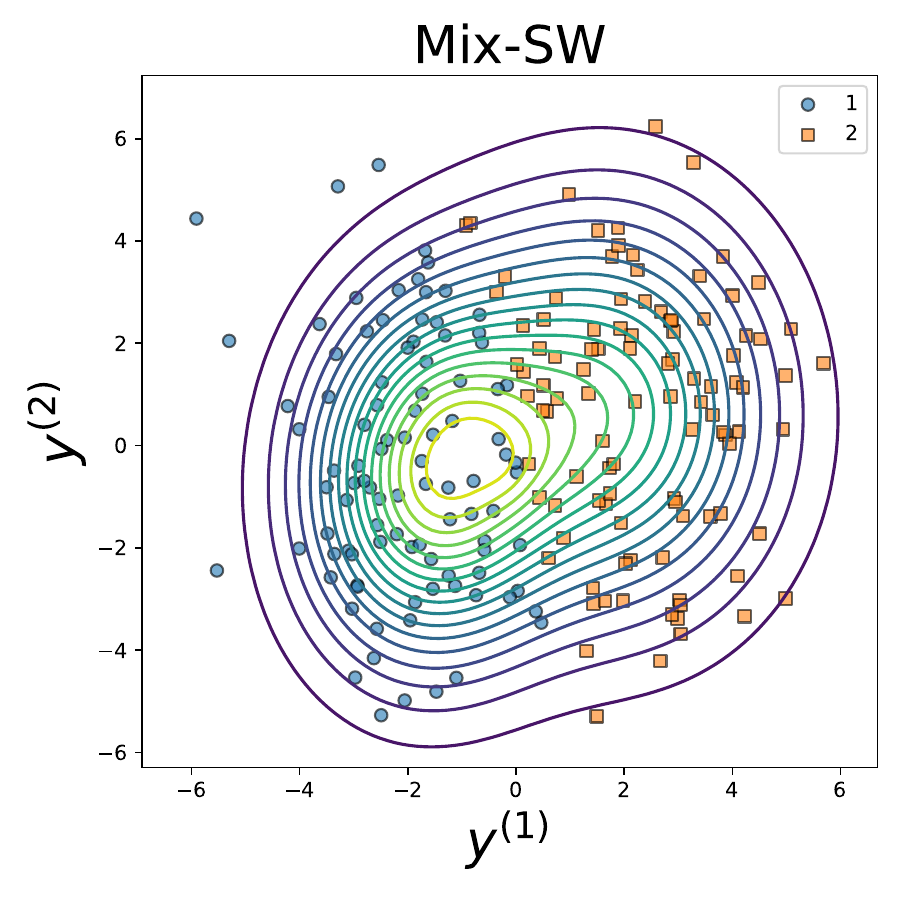} 
&\widgraph{0.25\textwidth}{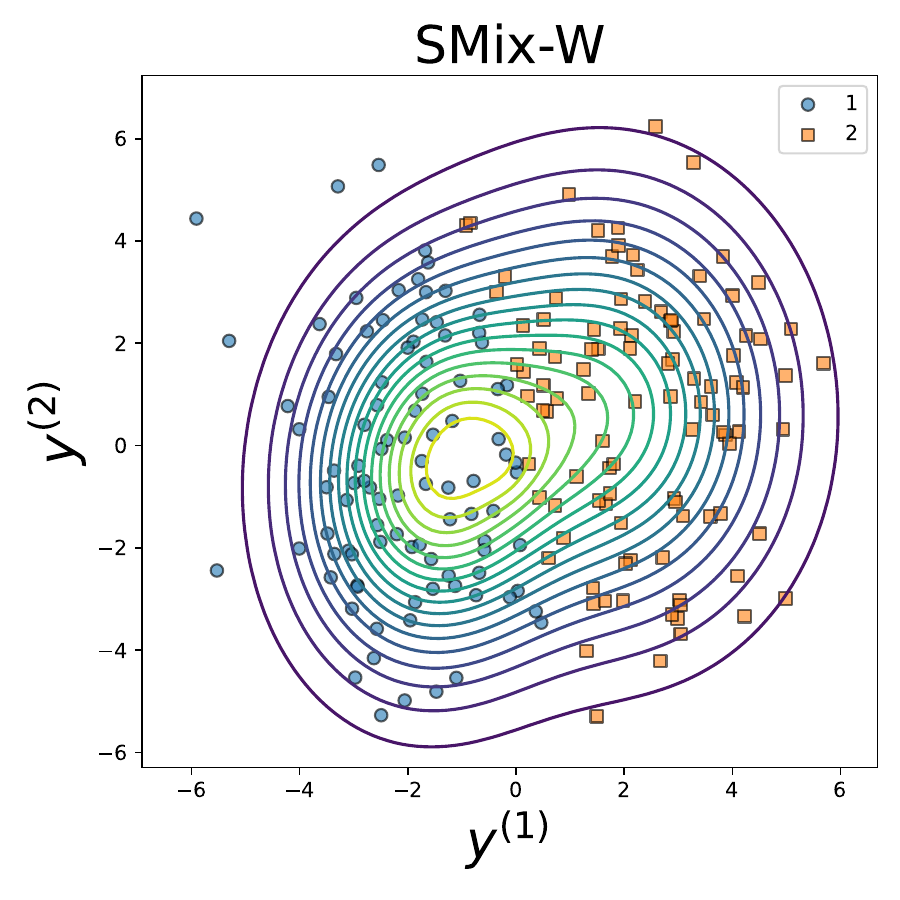}

  \end{tabular}
  \end{center}
  \vspace{-0.3 in}
  \caption{
  \footnotesize{{The figure shows the simulated data (dots), the true
    generating density (first plot, contours), the true cluster indices (first plot), the estimated density and
    cluster indices under different loss functions (other plots). For Binder, VI, and omARI loss, the estimated density is conditional on the partition. For SW, Mix-SW, and SMix-W, the contours show the estimated mixing measure and the Gaussian kernel}. 
}
} 
  \label{fig:simluated_data}
\end{figure}

\partitle{Visualization.}
Figure~\ref{fig:simluated_data} 
plots the simulated data, the true generating density, the true
cluster arrangements, and the density and estimated clusters obtained under
different loss functions in the two summarization approaches {from the first simulation repeatition}.
We evaluate the density on a \(100 \times 100\) grid, with
the range defined by $(\min y_i-1, \max y_i+1)$.
 For the first approach, starting with a point estimate of the partition,
we first determine  
\(\cbh\) (using Binder, VI, or omARI loss), and then evaluate
\(\EE[F \mid \cbh,y_{1:n}]\) by using 10 more iterations of the MCMC
simulation to update $F$ (freezing $\cbh$) {i.e., we evaluate $\mathbb{E}[F(x)\mid \cbh,y_{1:n}] \approx \frac{1}{10}\sum_{i=1}^{10} F_i(x)$  with $F_1,\ldots,F_{10} \sim p(F\mid \cbh,y_{1:n})$}. 

{
Table~\ref{tab:clustering_simulated_data} reports
the mean and the standard deviation of expected losses and
the relative loss, relative to the simulation truth, using Binder, VI, and omARI, from 25 repeated simulations. The expected losses reflect the quality of the point estimate with respect to the posterior while the relative losses reflect the quality of the point estimate in the frequentist sense.
We note that summarizing the random mixing measures yields comparable mean expected
Binder, mean expected VI, and mean expected omARI loss compared to the alternative approach of
summarizing the partition first despite the evaluation metric using the same losses.
In particular, summarizing the
random mixing measure always yields the second-best results for all clustering metrics. The proposed approach  yields a  smaller number of clusters than summarizing random partitions with Binder and omARI loss. All SW metrics behave similarly in terms of expected Binder  and expected omARI loss. SW yields a lower expected VI loss than Mix-SW and SMix-W. However, Mix-SW leads to the best relative Binder and  relative VI while SMix-W yield the best relative omARI loss. We
conclude that with respect to the reported point estimate $\cbh$
the approaches that start with a point estimate of the random mixing measure perform comparable to approaches that start with the random
partition. In addition, all SW variants seem to work similarly in terms of clustering performance.}

{
Table~\ref{tab:density_estimation_simulated_data} reports the means and the standard deviations of expected
distances and relative distances, relative to the simulation truth using total
variation distance and \(SW_2\) distance.
For the alternative approaches that first summarize the partitions (based on Binder, VI, or
omARI loss), the losses are averaged over 
10 Monte Carlo samples of \(F\) (freezing the point estimate $\cbh$) as discussed.
Naturally, the proposed approach of summarizing the mixing measures
first results in lower losses. Among SW variants, Mix-SW leads to the best total variation distances (both expected distance and relative distance). SMix-W leads to the best point estimate of the density in terms of both expected SW and relative SW distance.
}

{
Table~\ref{tab:mm_estimation_simulated_data} reports
the means and the standard deviations of expected losses and relative losses, relative to the simulation truth,
using \(SW_2\) distance, Mix-\(SW_2\) distance, and SMix-\(W_2\) distance from 25 repeat simulations.
Not surprisingly, inference under each distance performs best when used
for its intended evaluation.  
}

\begin{table}[!t]
    \centering
    \scalebox{0.68}{
    \begin{tabular}{|c|c|c|c|c|c|c|c|}
    \toprule
         &$k^*_n$&$\mathbb{E}[B(\hat{z},z)]$ & $B(\hat{z},z^\star) $ &$\mathbb{E}[VI(\hat{z},z)]$ & $VI(\hat{z},z^\star) $&$\mathbb{E}[omARI(\hat{z},z)]$ & $omARI(\hat{z},z^\star)$\\
         \midrule 
         Binder &$3.52 \pm 3.08$&$\mathbf{0.1615 \pm 0.0887}$&$0.5103 \pm 0.2112$&$0.7818 \pm 0.4968$&$1.9904 \pm 0.1501$&$0.6654 \pm 0.2978$&$0.8213 \pm 0.1672$\\
         VI &$1.52 \pm 0.59$&$0.1702 \pm 0.1039$&$0.5495 \pm 0.2119$ &$\mathbf{0.6976 \pm 0.4172}$&$\mathbf{1.9216 \pm 0.1343}$&$0.7070 \pm 0.3155$&$0.8442 \pm 0.1721$\\
         omARI&$5.76 \pm 2.26$ &$0.2170 \pm 0.0970$&$\mathbf{0.4362 \pm 0.1609}$&$1.0043 \pm 0.3267$&$2.0480 \pm 0.2323$&$\mathbf{0.6065 \pm 0.2385}$&$\mathbf{0.7783 \pm 0.1545}$\\
         \midrule
         SW&$1.76 \pm 0.78$&$0.1651 \pm 0.0909$&$0.5150 \pm 0.2047$&$0.7474 \pm 0.4558$&$1.9352 \pm 0.1352$ &$0.6687 \pm 0.2944$&$0.8225 \pm 0.1643$\\
         Mix-SW&$1.84 \pm 0.85$&$0.1661 \pm  0.0920$&$0.5057 \pm 0.2064$& $0.7684 \pm 0.4764$&$1.9342 \pm 0.1449$&$0.6657 \pm  0.2912$&$0.8140 \pm 0.1709$\\
         SMix-W&$2.04 \pm 1.02$& $0.1673 \pm 0.0933$&$0.5021 \pm 0.2067$&$0.7938 \pm 0.4916$&$1.9616 \pm 0.1416$&$0.6667 \pm 0.2868$&$0.8134 \pm 0.1701$\\
         \bottomrule
    \end{tabular}
    }
    \caption{{Simulated data: the table summarizes the estimated partitions under different loss functions. The columns from left to right are the number of unique clusters, expected Binder loss, relative Binder loss, expected VI loss, relative VI loss, expected omARI loss, and relative omARI loss. Lower losses are better.}}
    \label{tab:clustering_simulated_data}
\end{table}

\begin{table}[!t]
    \centering
    \scalebox{0.8}{
    \begin{tabular}{|c|c|c|c|c|}
    \toprule
         &$\mathbb{E}[TV(\hat{F},F))$&$TV(\hat{F},F^\star)$&$\mathbb{E}[SW_2(\hat{F},F))$&$SW_2(\hat{F},F^\star)$\\
         \midrule 
         Binder& $0.2363\pm0.0639$&$0.4262\pm0.0281$&$0.2845\pm0.0201$&$0.5807\pm0.0898$\\
         VI& $0.2267\pm0.0516$ &$0.4140\pm0.0291$&$0.2858\pm0.0196$&$0.5742\pm0.1034$\\
         omARI&$0.2650\pm 0.0566$&$0.4392\pm0.0303$&$0.2950\pm 0.0235$&$0.5777\pm 0.0955$ \\
         \midrule
         SW&$0.1947\pm 0.0569$&$0.4032\pm0.0424$&$0.2424\pm0.0384$&$0.5695\pm 0.1238$\\
         Mix-SW&$\mathbf{0.1907\pm0.0533}$&$\mathbf{0.4019\pm0.0368}$&$0.2318\pm 0.0283$&$0.5614 \pm 0.1116$\\
         SMix-W&$0.1955\pm0.0556$&$0.4060\pm0.0282$&$\mathbf{0.2282\pm0.0262}$&$\mathbf{0.5600\pm 0.1008}$\\
         \bottomrule
    \end{tabular}
    }
    
    \caption{{Simulated data: the table summarizes the estimated density under different loss functions. Columns from left to right are  expected Total Variation distance, Total Variation to the true density, expected $SW_2$ distance, and $SW_2$ distance to the true density. Lower distances are better.}}
    \label{tab:density_estimation_simulated_data}
\end{table}

\begin{table}[!t]
    \centering
    \scalebox{0.7}{
    \begin{tabular}{|c|c|c|c|c|c|c|}
    \toprule
         &$\mathbb{E}[SW_2(\Gh,G))$&$SW_2(\Gh,G^\star)$ &$\mathbb{E}[\text{Mix-}SW_2(\Gh,G))$&$\text{Mix-}SW_2(\Gh,G^\star)$ &$\mathbb{E}[\text{SMix-}W_2(\Gh,G))$&$\text{SMix-}W_2(\Gh,G^\star)$\\
         \midrule 
         SW& $\mathbf{0.8492\pm0.1881}$&$\mathbf{1.9826\pm0.5224}$&$0.5315\pm 0.1553$&$1.2406\pm0.2891$&$0.4591\pm0.1600$&$1.1762 \pm 0.2701$\\
         Mix-SW&$0.8882\pm0.2027$&$2.0051 \pm 0.4975$&$\mathbf{0.5183\pm0.1528}$&$\mathbf{1.2278\pm 0.2893}$&$0.4446\pm0.1574$&$1.1648\pm 0.2694$\\
         SMix-W&$0.8991\pm0.2128$&$2.0283\pm0.5401$&$0.5264\pm0.1567$&$1.2254\pm0.2950$&$\mathbf{0.4410\pm0.1563}$&$\mathbf{1.1574\pm0.2740}$\\
         \bottomrule
    \end{tabular}
    }
    
    \caption{{Simulated data: evaluation of  point estimates of the mixing measures $G$ under SW, Mix-SW, and SMix-W. The columns from left to right are  expected $SW_2$ distance, $SW_2$ distance to the simulation true $G^\star$, expected $\text{Mix-S}W_2$ distance, $\text{Mix-S}W_2$ distance $G^\star$, expected $\text{SMix-}W_2$ distance, and  $\text{SMix-}W_2$ distance to $G^\star$. Lower distances are better.}}
    \label{tab:mm_estimation_simulated_data}
\end{table}

{For summaries based on  $1000$ posterior MCMC samples of the mixing measures (truncation level $K=100$), SW and SMix-W take about $10$ minutes while Mix-SW takes about $40$ minutes on a Nvidia Geoforce RTX 3060 with Pytorch implementation. Summarizing random partition with the SALSO package (R implementation), takes about $2$ minutes for  $1000$ MCMC posterior samples of $200$ cluster indices using any of the three losses (Binder, VI, omARI). The reason that summarizing mixing measure is slower is  that the mixing measure contains more information than the cluster indices. The computation of summarizing the mixing measure depends on the truncation level $K$, the dimension $d$, but does not depend on the number of data points $n$. In contrast, summarizing the partition depends only on $n$. Therefore, with a very large $n$, summarizing the mixing measure might be more efficient. Once the point estimate of the mixing measure is evaluated, it takes only $1$ minutes to obtain the point estimate of the random partition and the density. In contrast, it takes around $5$ minutes to evaluate the density estimate with 10 more MCMC iterations starting from an estimated partition.}

\subsection{Old Faithful Geyser dataset}
\label{subsec:old_faithful}

\begin{figure}[!t]
\begin{center}
    \begin{tabular}{cc}
  \widgraph{0.25\textwidth}{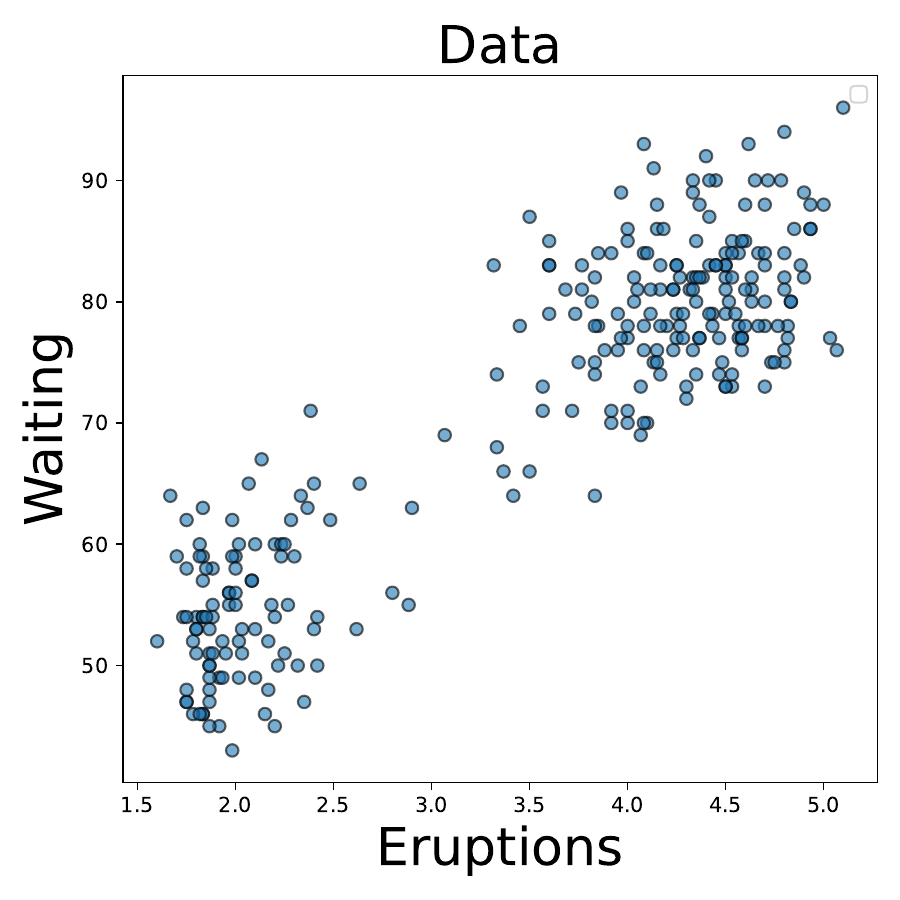} &\widgraph{0.25\textwidth}{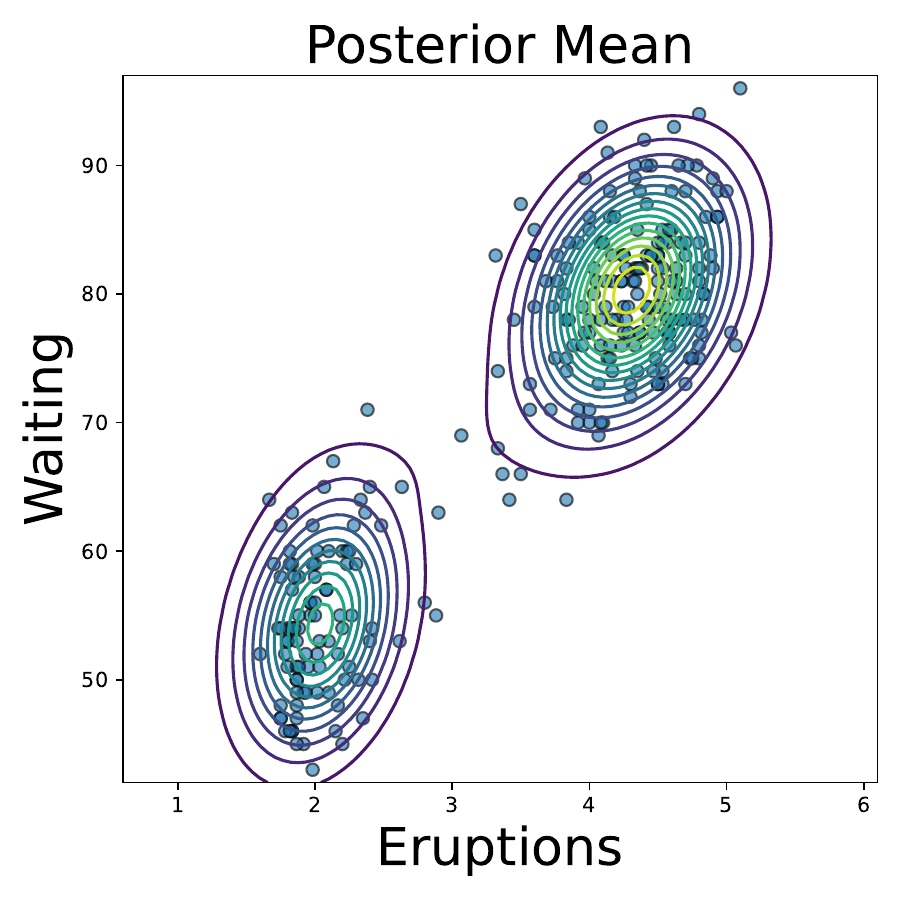}  

  \end{tabular}
  \begin{tabular}{ccc} 
\widgraph{0.25\textwidth}{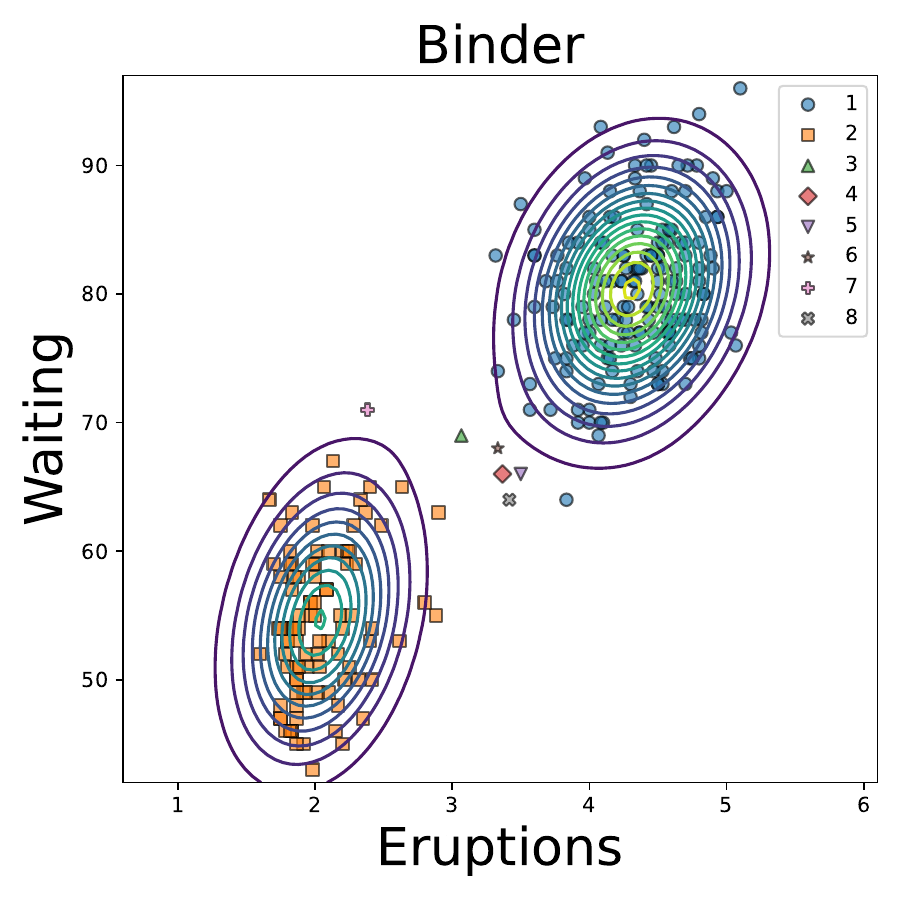} 
&\widgraph{0.25\textwidth}{images/faithful/vi_density.pdf} 
 &\widgraph{0.25\textwidth}{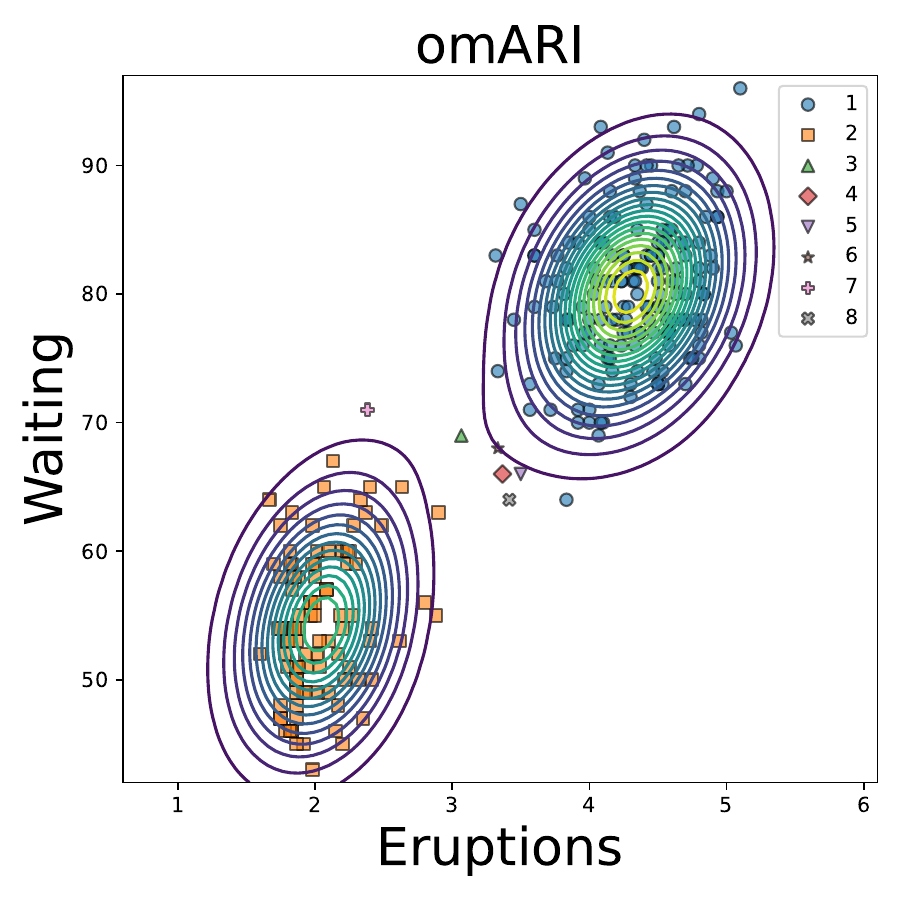} \\ \widgraph{0.25\textwidth}{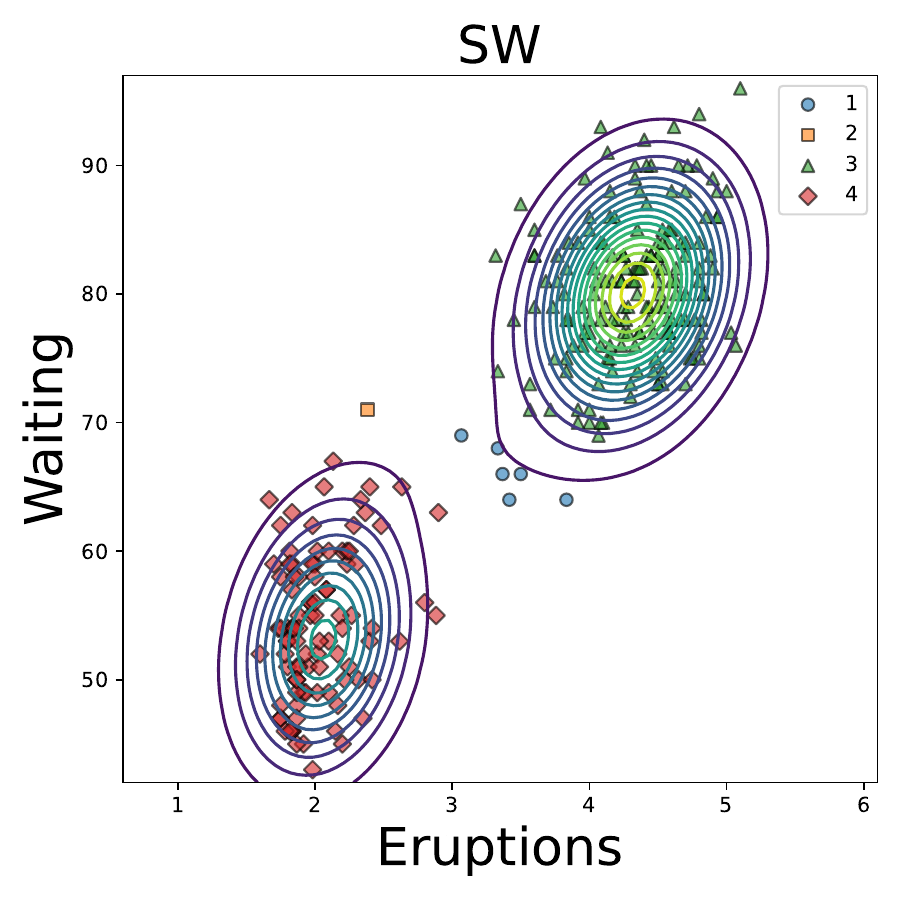}  
&\widgraph{0.25\textwidth}{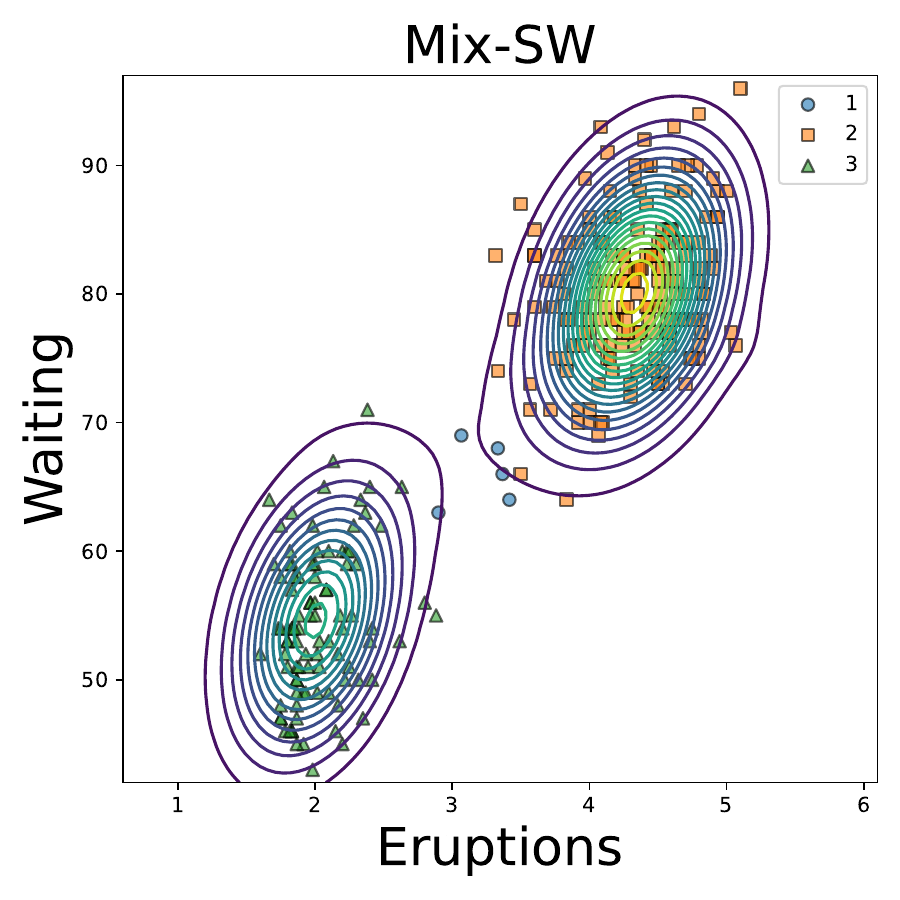} 
&\widgraph{0.25\textwidth}{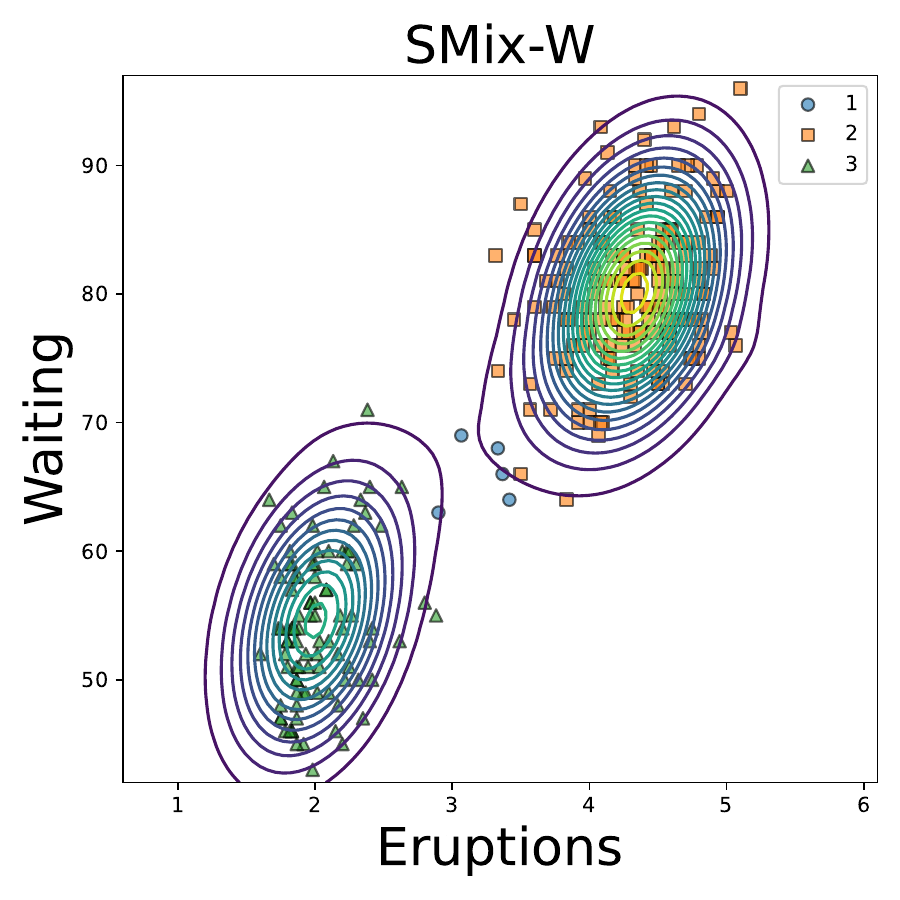}

  \end{tabular}
  \end{center}
  \vspace{-0.3in}
  \caption{
    \footnotesize{Old Faithful: same as Figure~\ref{fig:simluated_data}.
}
} 
  \label{fig:faithful}
\end{figure}

\partitle{Data and Inference.}
The Old Faithful geyser dataset contains 272 data samples in 2
dimensions. We run 10000 blocked Gibbs sampler iterations (9000
burn-in iterations) with the following hyperparameters
$\mu_0=(3,70),\Psi=diag((4,26)),\lambda=1,\nu=4,\alpha=1,K=100$. 
\partitle{Visualization.}
Figure~\ref{fig:faithful}
shows the data, along with the density estimate and estimated
partition obtained under different loss functions using the two
 approaches.
 
% \partitle{Clustering performance.} 
Table~\ref{tab:clustering_oldf} reports the expected losses
using Binder loss, VI, and omARI loss.
We note a similar overall pattern as in the simulation:
summarizing the random mixing measures yields comparable expected
Binder loss and expected omARI loss compared to partition-focused
approaches. 
In particular, summarizing the random mixing measures with
SW, Mix-SW, and SMix-W results in only about a \(3.38\%\) increase in
expected Binder loss compared to the best expected Binder loss, and
only about a \(3.37\%\) increase in expected omARI loss compared to
the best expected omARI loss. For VI loss, the proposed approach
results in an increase of about \(8.86\%\) in expected VI loss
compared to the best expected VI loss. 
%
% \partitle{Density estimation performance.}
Table~\ref{tab:density_estimation_oldf} 
reports the expected losses using total variation and \(SW_2\)
distance for the density estimation.
For the partition-focused approaches (using Binder, VI, or
omARI loss), the losses are averaged over 10 Monte Carlo samples of \(F\),
as under the previously described simulation study.
Summarizing the mixing
measure naturally leads to better density estimates than summarizing
partitions first. In particular, the best partition-focused method
reports a \(13.5\%\) higher expected total variation loss and \(19.23\%\)
higher expected \(SW\) loss compared to Mix-SW and
SMix-W. Furthermore, we find that SMix-W and Mix-SW compare
favorably to SW. 

\partitle{Mixing measure estimation performance.}
Finally, in Table~\ref{tab:mm_estimation_oldf}
we report the expected losses under \(SW_2\) distance, Mix-\(SW_2\) distance,
and SMix-\(W_2\) distance for estimating the mixing measure.
Again we {observe} similar patterns as in the simulation study.
Inference under each loss function performs best when used for
its intended evaluation. 

\begin{table}[!t]
    \centering
    \scalebox{0.8}{
    \begin{tabular}{|c|c|c|c|c|}
    \toprule
         &$k_n^\star$&$\mathbb{E}[B(\hat{z},z)]$ &$\mathbb{E}[VI(\hat{z},z)]$  &$\mathbb{E}[omARI(\hat{z},z)]$ \\
         \midrule 
         Binder &8&\textbf{0.0296}&0.2588&\textbf{0.0594} \\
         VI &3&0.0333&\textbf{0.2446}& 0.0667\\
         omARI&8&\textbf{0.0296}&0.2588&\textbf{0.0594}\\
         \midrule
         SW&4&0.0303&0.2602&0.0607\\
         Mix-SW&3&0.0306&0.2678&0.0614\\
         SMix-W&3&0.0306&0.2678&0.0614\\
         \bottomrule
    \end{tabular}
    }
    \caption{ Old
      Faithful: same as corresponding columns in Table~\ref{tab:clustering_simulated_data}. } 
    \label{tab:clustering_oldf}
\end{table}

\begin{table}[!t]
    \centering
    \scalebox{0.8}{
    \begin{tabular}{|c|c|c|c|}
    \toprule
         &$\mathbb{E}[TV(\hat{F},F))$&$\mathbb{E}[SW_2(\hat{F},F))$\\
         \midrule 
         Binder& 0.2128&1.0558\\
         VI& 0.2102&1.0293\\
         omARI&0.2128&1.0558\\
         \midrule
         SW&0.1901&0.8671\\
         Mix-SW&\textbf{0.1852}&0.7759\\
         SMix-W&\textbf{0.1852}&\textbf{0.7752}\\
         \bottomrule
    \end{tabular}
    }
    \caption{Old Faithful: same as Table~\ref{tab:density_estimation_simulated_data}  (columns 1 and 3).}
    \label{tab:density_estimation_oldf}
\end{table}

\begin{table}[!t]
    \centering
    \scalebox{0.8}{
    \begin{tabular}{|c|c|c|c|}
    \toprule
         &$\mathbb{E}[SW_2(\Gh,G))$&$\mathbb{E}[\text{Mix-}SW_2(\Gh,G))$&$\mathbb{E}[\text{SMix-}W_2(\Gh,G))$\\
         \midrule 
         SW&\textbf{3.1228}& 1.4876& 1.4831\\
         Mix-SW& 3.7931& \textbf{1.3186}& \textbf{1.3091} \\
         SMix-W& 3.7917& \textbf{1.3186}& \textbf{1.3091}\\
         \bottomrule
    \end{tabular}
    }
    
    \caption{Old Faithful: same as corresponding columns in Table~\ref{tab:mm_estimation_simulated_data}. }
    \label{tab:mm_estimation_oldf}
\end{table}
\section{Conclusion}
\label{sec:conc}
We present a new approach for {summarizing posterior inference} under  Bayesian
Nonparametric (BNP) mixture models. {The main difference to commonly used methods is that we start with a summary of}  the mixing measures. Our method minimizes
the posterior 
expected loss using a discrepancy between measures, utilizing the computationally scalable SW
distance. For Gaussian mixture models, we introduce two variants:
mixed sliced Wasserstein (Mix-SW) and sliced mixture Wasserstein
(SMix-W). Mix-SW uses generalized geodesic projection on the product
of the Euclidean manifold and the manifold of symmetric positive
definite matrices, providing a principled meaningful metric for comparing
Gaussian mixing measures. SMix-W leverages the linearity of Gaussian
mixtures for efficient projection. Empirical analyses show that our
summarization approach yields more accurate density estimates while
maintaining a good partition summary.  

% Future work will focus on enhancing the search algorithm and extending the method to more generalized mixture models.

Limitations of the proposed approach include the potential suboptimality of the reported
point estimate of the mixing measures, as the solution is limited to
the available posterior Monte Carlo samples. In addition,
the conventional sliced
Wasserstein distance might not be optimal for all mixture models;
therefore, different variants should be designed to exploit specific
geometry, as we do for Gaussian mixtures. 

{The proposed framework addresses the problem of summarizing
posterior inference on a random partition when it is being used, leaving the judgement about appropriate
use of the chosen model to the investigators. In particular, it has been observed that  clustering of high-dimensional data might be challenging~\citep{chandra2023escaping}.} 

\clearpage
\bigskip
{
\begin{center} {\large\bf SUPPLEMENTARY MATERIALS}
\end{center}
The Supplementary Materials include Appendices and codes to reproduce empirical parts of the paper. 
\begin{description}
\item[Appendices:] Appendices are in the file ``appendix.pdf" which contains technical proofs and additional discussion.
\item[Python code,  R code, and data:] The file ``code.zip” contains Python code, R code, and data to reproduce all empirical parts of the paper. We refer the reader to the ``README.md” file inside the zip archive for detailed instructions.
\end{description}
}
\bibliographystyle{jasa3}

\bibliography{bibtex}

\clearpage
\appendix
\section{Appendices}
\label{sec:appendix}

\subsection{Acronyms}
\label{subsec:acronyms}
{We summarize the a list of acronyms used in the paper as follow:
\begin{itemize}
    \item BNP: Bayesian Nonparametric.
    \item SW: sliced Wasserstein.
    \item Mix-SW: mixed sliced Wasserstein.
    \item SMix-W: sliced mixture Wasserstein.
    \item MAP: maximum a posteriori.
    \item MCMC: Markov chain Monte Carlo.
    \item VI: variational information.
    \item omARI: one minus adjusted Rand index.
\end{itemize}
}
\subsection{Review on Riemannian Manifolds}
\label{subsec:manifold_review}
A Riemannian manifold ($\mathcal{M},G$) of dimension $d$ is a space that behaves locally
as a linear space diffeomorphic to $\Re^d$, named a tangent space. For any $x \in \mathcal{M}$, the associated tangent space is defined as $T_x \mathcal{M}$ which supports an inner product $\langle\cdot,\cdot \rangle_x: T_x \mathcal{M} \times T_x \mathcal{M} \to \Re$ i.e., $\langle u,v \rangle_x = u^\top G(x) v $. The joint space of the manifold and the tangent space is called the tangent bundle $T\mathcal{M}=\{x \in \mathcal{M},v \in T_x\mathcal{M}\}$.

\paragraph{Geodesics.} Given two points $x,y \in \mathcal{M}$, the smooth curve $\gamma:[0,1] \to \mathcal{M}$ such as $\gamma(0)=x,\gamma(1)=y$ is called geodesic if it minimizes the length:
\begin{align*}
    \mathcal{L}(\gamma) =  \int_{0}^1 \sqrt{\langle \gamma'(t) ,\gamma'(t)\rangle_{\gamma(t)}} \mathrm{d}t,
\end{align*}
where $\gamma'(t)$ is the derivative of the curve $\gamma(t)$ with respective to $t$, which belongs to the tangent space $T_{\gamma(t)} \mathcal{M}$ for any $t \in [0,1]$.
% \mynote{K: is it worth to say how to interpret/define $\gamma'(t)$
%   here? Is it defined by the derivative on the tangent space?}
 The length of the geodesic line is the geodesic distance.
\begin{align*}
    c(x,y) = \inf_{\gamma|\gamma(0)=x,\gamma(1)=y} \mathcal{L}(\gamma).
\end{align*}

\paragraph{Exponential Map.} Let $x\in \mathcal{M}$, for any $v \in T_x\mathcal{M}$, there exists a unique geodesic $\gamma$ with $\gamma(0)=x$ and $\gamma'(0)=v$, denoted as $\gamma_{x,v}$. The exponential map $\exp: T\mathcal{M} \to \mathcal{M}$ maps $v \in T_x \mathcal{M} $ back to the manifold at the point reached by the
geodesic $\gamma(1)$. In particular, we have the following definition of the exponential map:
\begin{align*}
    \forall (x,v) \in T\mathcal{M}, \exp_x(v)= \gamma_{x,v} (1)
\end{align*}

\subsection{Mixed Sliced Wasserstein}
\label{subsec:appendix:mixSW}
\paragraph{Geodesic Projection}
Let \( \gamma \) be a curve on the manifold \( \mathcal{M} \), and denote \( \mathcal{A} \) as the set of all points belonging to that curve. The projection of a point \( x \in \mathcal{M} \) onto the curve \( \gamma \) is defined as: $
    \tilde{P}_\gamma (x) = \text{arg}\min_{y \in \mathcal{A} } c(x, y),$
where \( c \) is {a} geodesic distance. If we constrain \( \gamma \) to be a curve that passes through the origin (denoted as \( o \)) with unit velocity \( v \) (i.e., \( \langle v, v \rangle_o = 1 \)), then we have \( \Gamma = \{ \exp_o(tv) \mid t \in \Re \} \), where \( \exp_o(\cdot) \) is the exponential map at the origin. The coordinates of the projection can be determined by solving:
\begin{align}
    P_{\gamma_{(o,v)}}(x) := P_v(x) = \text{arg}\min_{t \in \Re} 
  c(x, \exp_o(tv)).
\end{align}

% \paragraph{Manifold of symmetric positive definite matrices $S_d^{++}(\Re)$.} 

\paragraph{Product Manifold of \( \Re^d \times S_d^{++}(\Re) \).} From~\citet{pennec2019riemannian}, the origin of \( S_d^{++}(\Re) \) is the identity matrix \( I \), and the tangent space is the space of all symmetric matrices. The exponential map is given by $
    \exp_I(A) = \exp(A) = \sum_{n=0}^\infty \frac{A^n}{n!}.$ While there are multiple geodesic distances on \( S_d^{++}(\Re) \), we focus on the Log-Euclidean metric defined as:
$
    c_{LE}(\Sig_1, \Sig_2) = \|\log \Sig_1 - \log \Sig_2\|_F,
$
where \( \log X = A \) if \( \exp(A) = X \). Since \( X \) is a symmetric positive definite matrix, we can use spectral decomposition: \( X = Q \Lambda Q^T \), which gives us $
    \log X = Q \, \text{diag}(\log \lambda_1, \ldots, \log \lambda_d) \, Q^T.$ The origin of the manifold \( \Re^d \times S_d^{++}(\Re) \) is \( o = (0, I) \), where \( 0 \) is the \( d \)-dimensional zero vector. The exponential map in this manifold is defined as $
    \exp_o((\mu, \Sig)) = (\mu, \exp(\Sig))$.
A geodesic distance of the product manifolds is defined as follows~\citep{gu2019learning}: $ 
    c((\mu_1, \Sig_1), (\mu_2, \Sig_2)) = \sqrt{\|\mu_1 - \mu_2\|_2^2 + c_{LE}(\Sig_1, \Sig_2)^2}.
$

\subsection{Sliced Mixture Wasserstein}
\label{subsec:appendix:smix}

\paragraph{Mixture Wasserstein distance.} Given two discrete
% \mynote{K: do they need to be finite? In (17) you do the $\inf$ over a
%   subset of finite Gaussian mixtures in $2d$ dims. Which can only work
%   if $F_1$, $F_2$ are finite?}
measures $G_1$ and $G_2$ belonging to $\PP(\Re^d \times S_d^{++}(\Re))$, and 
a Gaussian kernel $f(x \mid \mu, \Sig)$, 
we define $F_1$ and $F_2$ as the corresponding mixtures of
Gaussian measures, i.e., $F_1 = f * G_1$ and $F_2 = f * G_2$, where
$*$ denotes the standard convolution operation. The Mixture
Wasserstein (MW) distance~\citep{delon2020wasserstein} is defined as: 
\begin{align}
    \MW_2^2 (F_1, F_2) = \inf_{\pi \in \Pi(F_1, F_2)\, \cap\, \GMM_{2d}(\infty)} \int_{\Re^d \times \Re^d} \| x - y \|_2^2 \, \mathrm{d}\pi(x, y),
\end{align}
where $\GMM_{2d}(\infty)$ denotes the set of all finite Gaussian mixture distributions in $2d$ dimensions. Compared to Wasserstein distance, Mixture Wasserstein consider only couplings that are finite Gaussian mixtures. When $G_1 = \sum_{i=1}^{K_1} \alpha_i \delta_{(\mu_{1i}, \Sig_{1i})}$ and $G_2 = \sum_{j=1}^{K_2} \beta_j \delta_{(\mu_{2j}, \Sig_{2j})}$, the Mixture Wasserstein distance simplifies to:
$
    \min_{\eta \in \Gamma(\alpha, \beta)} \sum_{i=1}^{K_1} \sum_{j=1}^{K_2} \eta_{ij} W_2^2\left(\mathcal{N}(\mu_{1i}, \Sig_{1i}), \mathcal{N}(\mu_{2j}, \Sig_{2j})\right).
$
Using the closed-form expression of the Wasserstein-2 distance between two Gaussian distributions, we can rewrite this as:
$$
    \min_{\eta \in \Gamma(\alpha, \beta)} \sum_{i=1}^{K_1} \sum_{j=1}^{K_2} \eta_{ij} \left( \|\mu_{1i} - \mu_{2j}\|_2^2 + \mathrm{Tr}(\Sig_{1i}) + \mathrm{Tr}(\Sig_{2j}) \right. \\ \left.- \mathrm{Tr}\left((\Sig_{1i}^{1/2} \Sig_{2j} \Sig_{1i}^{1/2})^{1/2}\right) \right).
$$

\paragraph{One-dimensional Mixture Wasserstein distance.}
In preparation of the upcoming definition of sliced MW by
comparing one-dimensional projections of mixtures of normals, we note
the special case of one-dimensional MW distance. 
When $G_1 = \sum_{i=1}^{K_1} \alpha_i \delta_{(\mu_{1i}, \sig_{1i}^2)}$ and $G_2
= \sum_{j=1}^{K_2} \beta_j \delta_{(\mu_{2j}, \sig_{2j}^2)}$ are
one-dimensional mixtures of Gaussians, we have: 
\begin{multline}
    \MW_2^2(F_1, F_2) = \min_{\gamma \in \Gamma(\alpha, \beta)}
                        \sum_{i=1}^{K_1} \sum_{j=1}^{K_2} \gamma_{i,j}
                        \left((\mu_{1i} - \mu_{2j})^2 + (\sig_{1i} -
                          \sig_{2j})^2\right)  \\
    = W_2^2\left((\mathrm{Id},
    \sqrt{\cdot})\sharp G_1, (\mathrm{Id}, \sqrt{\cdot})\sharp G_2\right),
\label{MW1D}                        
\end{multline}
which implies that the Mixture Wasserstein distance between one-dimensional Gaussian mixtures behaves like a two-dimensional Wasserstein-2 distance on the mixing measures, with a square root scaling applied to the variances. It is important to note that the one-dimensional MW distance does not offer computational advantages, as it is not equivalent to a one-dimensional Wasserstein distance.

\paragraph{Sliced Mixture Wasserstein.} After applying linear
projection to the mixture Gaussians (or $P'_v$ on the Gaussian mixing
measure), we can use  one-dimensional MW in \eqref{MW1D}  to
compare them i.e.,
\begin{align}  
\label{eq:SMW}
    \mathbb{E}_{v \sim \mathcal{U}(\mathbb{S}^{d-1})}[\MW_2^2( P_v\sharp F_1, P_v\sharp F_2)]=\mathbb{E}_{v \sim \mathcal{U}(\mathbb{S}^{d-1})}[W_2^2((Id, \sqrt)\sharp P_v'\sharp G_1, (Id,\sqrt)\sharp P_v'\sharp G_2)].
\end{align}
However, MW does not have a closed-form expression, as discussed. Since MW is equivalent to the Wasserstein-2 distance between mixing measures, we can replace it with the SW distance to achieve computational benefits, as SW is equivalent to the Wasserstein distance under a mild assumption~\citep{bonnotte2013unidimensional} {i.e., atoms of two measures belong to a ball of size $R>0$}. This replacement leads to a novel variant of SW distance for mixtures of Gaussians and their mixing measures.

We can rewrite SMix-W in Definition~\ref{def:smixW} as:
$$
\text{SMix-}W_2^2(G_1,G_2) = \mathbb{E}_{v \sim
  \mathcal{U}(\mathbb{S}^{d-1})}\left[SW_2^2\left((\text{Id}, \log
    \circ \sqrt)\sharp P_v' \sharp G_1, \, (\text{Id}, \log \circ
    \sqrt)\sharp P_v' \sharp G_2\right)\right]
$$
which replaces MW of the mixtures in~\eqref{eq:SMW} with the SW of the mixing measures, incorporating a logarithmic transformation to adjust the standard deviation as a geodesic projection. Compared to SW and Mix-SW, the projection space of SMix-W is smaller, specifically \(\mathbb{S}^{d-1} \times \mathbb{S}\), as it utilizes a single projecting direction \(v\) for both the mean \(\mu\) and the covariance matrix \(\Sig\).
\begin{algorithm}[!t]
\caption{Summarizing the posterior of mixing measure}
\begin{algorithmic}
\label{alg:summarization}
\STATE \textbf{Input:} Posterior samples of the mixing measure $G_1,\ldots,G_M$ ($M\geq 2$), and discrepancy $\mathcal{D}$ (assumed to be symmetric).
\STATE Initialize $C \in \Re^{M \times M}$
  \FOR{$i=1$ to $M$}
  \FOR{$j=i+1$ to $M$} 
  \STATE $C_{ij}=C_{ji}=\mathcal{D}(G_i,G_j)$ 
  \ENDFOR
  \ENDFOR
  \STATE Find $i^\star = \argmin_{i \in \{1,\ldots,M\}} \sum_{j=1}^M C_{ij}$
 \STATE \textbf{Return:}   $G_{i^\star}$.
\end{algorithmic}
\end{algorithm}
\subsection{{Algorithms}}
\label{subsec:appendix:algorithms}

{We present the pseudo algorithm for the proposed posterior summarization framework in Algorithm~\ref{alg:summarization}. Moreover, we also give pseudo algorithms for vectorized SW, Mix-SW, and SMix-W in Algorithm~\ref{alg:SW}, Algorithm~\ref{alg:MixSW}, and Algorithm~\ref{alg:SMixW} in turn.}

\begin{algorithm}[!t]
\caption{Computational algorithm of vectorized sliced Wasserstein}
\begin{algorithmic}
\label{alg:SW}
\STATE \textbf{Input:} Probability measures $G_1=\sum_{i=1}^{K_1} \alpha_i \delta_{(\mu_{1i},\Sigma_{1i})}$ and $G_2=\sum_{j=1}^{K_2} \beta_j \delta_{(\mu_{2j},\Sigma_{2j})}$, $p\geq 1$,  function $V(\theta)$, and the number of projections $L$.
    
  \STATE Vectorize $V\sharp G_1 =\sum_{i=1}^{K_1} \alpha_i \delta_{V(\mu_{1i},\Sigma_{1i})}$ and $V\sharp G_2 =\sum_{j=1}^{K_2} \beta_j \delta_{V(\mu_{2j},\Sigma_{2j})}$
  \FOR{$l=1$ to $L$}
  \STATE Sample $v_l \sim \mathcal{U}(\mathbb{S}^{d(d+1)-1})$
  \STATE Compute $P_{v_l}\sharp V \sharp G_1 =  \sum_{i=1}^{K_1} \alpha_i \delta_{v_l^\top V(\mu_{1i},\Sigma_{1i})}$ and $v_l\sharp V\sharp G_2 =\sum_{j=1}^{K_2} \beta_j \delta_{v_l^\top V(\mu_{2j},\Sigma_{2j})}$
  \STATE Compute $\text{W}^p_p(P_{v_l}\sharp V \sharp G_1,P_{v_l}\sharp V \sharp G_2)$ as in~\eqref{eq:1DW}.
  \ENDFOR
  \STATE \textbf{Return:}  $\widehat{SW}_p(G_1,G_2;L) = \left(\frac{1}{L}\sum_{l=1}^L \text{W}_p^p(P_{v_l}\sharp V \sharp G_1,P_{v_l}\sharp V \sharp G_2)\right)^{\frac{1}{p}}$.
\end{algorithmic}
\end{algorithm}

\begin{algorithm}[!t]
\caption{Computational algorithm of mixed sliced Wasserstein (Mix-SW)}
\begin{algorithmic}
\label{alg:MixSW}
\STATE \textbf{Input:} Probability measures $G_1=\sum_{i=1}^{K_1} \alpha_i \delta_{(\mu_{1i},\Sigma_{1i})}$ and $G_2=\sum_{j=1}^{K_2} \beta_j \delta_{(\mu_{2j},\Sigma_{2j})}$, $p\geq 1$, and the number of projections $L$.
    
  \FOR{$l=1$ to $L$}
  \STATE Sample $v_l \sim \mathcal{U}(\mathbb{S}^{d-1}), A_l \sim \mathcal{U}(S_d(\Re)), w_l \sim \mathcal{U}(\mathbb{S})$.
 \STATE Set $V_{w,l} = (w_l,v_l,A_l)$.
  \STATE Compute $P_{V_{w,l}} \sharp G_1 =  \sum_{i=1}^{K_1} \alpha_i \delta_{P_{V_{w,l}}(\mu_{1i},\Sigma_{1i})}$ and $P_{V_{w,l}} \sharp G_2 =  \sum_{j=1}^{K_2} \beta_j \delta_{P_{V_{w,l}}(\mu_{2j},\Sigma_{2j})}$ as in~\eqref{prop:generalized_geodesic_projection}.
  \STATE Compute $\text{W}^p_p(P_{V_{w,l}} \sharp G_1,P_{V_{w,l}} \sharp G_2)$ as in~\eqref{eq:1DW}.
  \ENDFOR
  \STATE \textbf{Return:}  $\widehat{\text{Mix-}SW}_p (G_1, G_2) = \left(\frac{1}{L} \sum_{l=1}^L W_p^p(P_{V_{w,l}} \sharp G_1, P_{V_{w,l}} \sharp G_2)\right)^{\frac{1}{p}}$.
\end{algorithmic}
\end{algorithm}

\begin{algorithm}[!t]
\caption{Computational algorithm of sliced mixture Wasserstein (SMix-W)}
\begin{algorithmic}
\label{alg:SMixW}
\STATE \textbf{Input:} Probability measures $G_1=\sum_{i=1}^{K_1} \alpha_i \delta_{(\mu_{1i},\Sigma_{1i})}$ and $G_2=\sum_{j=1}^{K_2} \beta_j \delta_{(\mu_{2j},\Sigma_{2j})}$, $p\geq 1$, and the number of projections $L$.
    
  \FOR{$l=1$ to $L$}
  \STATE Sample $v_l \sim \mathcal{U}(\mathbb{S}^{d-1}), w_l \sim \mathcal{U}(\mathbb{S})$.
  \STATE Compute $P_{v_l,w_l} \sharp G_1 =  \sum_{i=1}^{K_1} \alpha_i \delta_{P_{v_l,w_l}(\mu_{1i},\Sigma_{1i})}$ and $P_{v_l,w_l} \sharp G_2 =  \sum_{j=1}^{K_2} \beta_j \delta_{P_{v_l,w_l}(\mu_{2j},\Sigma_{2j})}$ as in~\eqref{prop:generalized_geodesic_projection}.
  \STATE Compute $\text{W}^p_p(P_{v_l,w_l} \sharp G_1,P_{v_l,w_l}\sharp G_2)$ as in~\eqref{eq:1DW}.
  \ENDFOR
  \STATE \textbf{Return:}  $\widehat{\text{SMix-}W}_p (G_1, G_2) = \left(\frac{1}{L} \sum_{l=1}^L W_p^p(P_{v_l,w_l} \sharp G_1, P_{v_l,w_l} \sharp G_2)\right)^{\frac{1}{p}}$.
\end{algorithmic}
\end{algorithm}

\subsection{Proof of Proposition~\ref{prop:generalized_geodesic_projection}}
\label{subsec:proof:prop:generalized_geodesic_projection}
By Definition~\ref{def:generalized_geodesic_projection}, we have:
\begin{align*}
    P_{V_w}((\mu,\Sig))&= \text{arg}\min_{t \in \Re }c(x,\exp_0(tV_w)) = \text{arg}\min_{t \in \Re }c^2(x,\exp_0(tV_w))  \\
&= \text{arg}\min_{t \in \Re } \|\mu-tw_1v\|_2^2 + \|\log \Sig -\log \exp(tw_2A)\|_F \\
        &= \text{arg}\min_{t \in \Re } \|\mu-tw_1v\|_2^2 + \|\log \Sig -tw_2A\|_F \\
        % &= \text{arg}\min_{t \in \Re } \|\mu\|_2^2 +t^2w_1^2\|v\|_2^2 -2tw_1 \langle \mu,v\rangle + t^2w_2^2 Trace(A^2)+Trace((log\Sig)^2)  -2tw_2Trace(V\log \Sig) \\
        &= \text{arg}\min_{t \in \Re } \|\mu\|_2^2 +w_1^2t^2 -2w_1t \langle \mu,v\rangle + w_2^2 t^2+Trace((log\Sig)^2)  -2w_2tTrace(A\log \Sig) \\
        &= \text{arg}\min_{t \in \Re } \|\mu\|_2^2 +t^2 -2w_1t \langle \mu,v\rangle +Trace((log\Sig)^2)  -2w_2tTrace(A\log \Sig) \\
        &:= \text{arg}\min_{t \in \Re } f(t)
\end{align*}
Taking the derivative $\frac{d}{dt} f(t) =2t - 2w_1 \langle \mu,v\rangle - 2w_2Trace(A\log \Sig)$, then set it to 0. We obtain:
$
    P_{V_w}(\theta) =t^\star = w_1 \langle \mu,v\rangle+ w_2Trace(A\log \Sig),
$
which completes the proof.

\subsection{Proof of Proposition~\ref{prop:boundness}}
\label{subsec:proof:prop:boundness}

     Since $P_{V_w}((\mu,\Sig)) =  w_1 \langle \mu,v\rangle+ w_2Trace(A\log \Sig)$ is a Borel meansurable, using Lemma 6 in ~\citep{paty2019subspace}, we have:
\begin{align*}
   & W_p^p(P_{V_w} \sharp G_1, P_{V_w}\sharp G_2)]  = \inf_{\pi_{V_w} \in \Pi(P_{V_w} \sharp G_1, P_{V_w}\sharp G_2)} \int_{\Re\times \Re} |x-y|^p \mathrm{d}\pi_{V_w}(x,y) \\
    &= \inf_{\pi \in \Pi( G_1, G_2)} \int_{\Re^d \times S_d^{++}(\Re) \times \Re^d \times S_d^{++}(\Re)} |P_{V_w}(\mu_1,\Sig_1)-P_{V_w}(\mu_2,\Sig_2)|^p \mathrm{d}\pi((\mu_1,\Sig_1),(\mu_2,\Sig_2)) \\
    &\leq \inf_{\pi \in \Pi( G_1, G_2)} \int_{\Re^d \times S_d^{++}(\Re) \times \Re^d \times S_d^{++}(\Re)} (|P_{V_w}(\mu_1,\Sig_1)-P_{V_w}(\mu_0,\Sig_0)|  \\& \quad + |P_{V_w}(\mu_0,\Sig_0)-P_{V_w}(\mu_2,\Sig_2)|)^p \mathrm{d}\pi((\mu_1,\Sig_1),(\mu_2,\Sig_2)),
\end{align*}
where the inequality is due to the triangle inequality of the $\mathbb{L}_1$ norm. Using the Minkowski's inequality, we have:
\begin{align*}
 W_p^p(P_{V_w} \sharp G_1, P_{V_w}\sharp G_2)] 
    &\leq \inf_{\pi \in \Pi( G_1, G_2))} \int_{\Re^d \times S_d^{++}(\Re) \times \Re^d \times S_d^{++}(\Re)} 2^{p-1} \left(|P_{V_w}(\mu_1,\Sig_1)-P_{V_w}(\mu_0,\Sig_0)|^p\right. \\ &\quad \left.+|P_{V_w}(\mu_0,\Sig_0)-P_{V_w}(\mu_2,\Sig_2)|^p \right) \mathrm{d}\pi_{V_w}\mathrm{d}\pi((\mu_1,\Sig_1),(\mu_2,\Sig_2)) \\
    &= 2^{p-1} \left(\int_{\Re^d \times S_d^{++}(\Re)}|P_{V_w}(\mu_1,\Sig_1)-P_{V_w}(\mu_0,\Sig_0)|^p  \mathrm{d}G_1(\mu_1,\Sig_1)  \right.\\
    &\quad +\left. \int_{\Re^d \times S_d^{++}(\Re)}|P_{V_w}(\mu_0,\Sig_0)-P_{V_w}(\mu_2,\Sig_2)|^p  \mathrm{d}G_2(\mu_2,\Sig_2) \right).
\end{align*}
Moreover, from the Cauchy–Schwarz's inequality, we have:
\begin{align*}
    &|P_{V_w}(\mu_1,\Sig_1)-P_{V_w}(\mu_0,\Sig_0)| \\
    &=| w_1 \langle \mu_1,v\rangle+ w_2Trace(A\log \Sig_1) - w_1 \langle \mu_0,v\rangle- w_2Trace(A\log \Sig_0) | \\
    &=|w_1 \langle \mu_1-\mu_0,v\rangle  +w_2(Trace(A\log \Sig_1)-Trace(A\log \Sig_0)) | \\
    &\leq \sqrt{w_1^2+w_2^2} \sqrt{\langle \mu_1-\mu_0,v\rangle^2 +(Trace(A (\log \Sig_1-\log \Sig_0))^2 } \\
    &= \sqrt{\langle \mu_1-\mu_0,v\rangle^2 +(Trace(A (\log \Sig_1-\log \Sig_0))^2 } \\
    &\leq \sqrt{\|v\|_2^2 \|\mu_1-\mu_0\|_2^2 +\|A\|_F^2 \|\log \Sig_1-\log \Sig_0\|_F^2} \\
    &= \sqrt{ \|\mu_1-\mu_0\|_2^2 + \|\log \Sig_1-\log \Sig_0\|_F^2 } =c((\mu_1,\Sig_1),(\mu_0,\Sig_0)).
\end{align*}
From the assumption, we get:
\begin{align*}
     W_p^p(P_{V_w} \sharp G_1, P_{V_w}\sharp G_2)] &\leq  2^{p-1} \left(\int_{\Re^d \times S_d^{++}(\Re)}c((\mu_1,\Sig_1),(\mu_0,\Sig_0))^p  \mathrm{d}G_1(\mu_1,\Sig_1)  \right.\\
    &\quad +\left. \int_{\Re^d \times S_d^{++}(\Re)}c((\mu_0,\Sig_0),(\mu_2,\Sig_2))^p \mathrm{d}G_2(\mu_2,\Sig_2) \right) < \infty,
\end{align*}
which completes the proof.

\subsection{Proof of Theorem~\ref{theorem:mixSW_metric}}
\label{subsec:proof:theorem:mixSW_metric}

\paragraph{Symmetry and Non-negativity.} The symmetry and non-negativity of Mix-SW follows directly the symmetry and non-negativity of the Wasserstein distance~\citep{peyre2020computational} since it is the expectation of projected Wasserstein distance.

\paragraph{Triangle Inequality.} Let consider three measure $G_1,G_2,G_3$, using the triangle inequality of Wasserstein distance, we have:
\begin{align*}
     \text{Mix-}SW_p(G_1,G_2) &= \left(\mathbb{E}_{(w,v,A)\sim \mathcal{U}(\mathbb{S}) \otimes \mathcal{U}(\mathbb{S}^{d-1}) \otimes \mathcal{U}(S_d(\Re)) } [W_p^p(P_{V_w} \sharp G_1, P_{V_w}\sharp G_2)]|\right)^{\frac{1}{p}} \\
&\leq \left(\mathbb{E}_{(w,v,A)\sim \mathcal{U}(\mathbb{S}) \otimes \mathcal{U}(\mathbb{S}^{d-1}) \otimes \mathcal{U}(S_d(\Re)) } [(W_p(P_{V_w} \sharp G_1, P_{V_w}\sharp G_3) +W_p(P_{V_w} \sharp G_3, P_{V_w}\sharp G_2))^p ] \right)^{\frac{1}{p}}.
\end{align*}
Using the Minkowski's inequality, we get:
\begin{align*}
    \text{Mix-}SW_p(G_1,G_2) &\leq \left(\mathbb{E}_{(w,v,A)\sim \mathcal{U}(\mathbb{S}) \otimes \mathcal{U}(\mathbb{S}^{d-1}) \otimes \mathcal{U}(S_d(\Re)) } [W_p^p(P_{V_w} \sharp G_1, P_{V_w}\sharp G_3)]|\right)^{\frac{1}{p}} \\
    &+ \left(\mathbb{E}_{(w,v,A)\sim \mathcal{U}(\mathbb{S}) \otimes \mathcal{U}(\mathbb{S}^{d-1}) \otimes \mathcal{U}(S_d(\Re)) } [W_p^p(P_{V_w} \sharp G_3, P_{V_w}\sharp G_2)]|\right)^{\frac{1}{p}} \\
    &=  \text{Mix-}SW_p(G_1,G_3) + \text{Mix-}SW_p(G_3,G_2),
\end{align*}
which completes the proof of triangle inequality.

\paragraph{Identity of indiscernibles.}  When $G_1=G_2$, we have directly $\text{Mix-}SW_p^p(G_1,G_2)=0$. We now prove that if $\text{Mix-}SW_p^p(G_1,G_2)=0$, we get $G_1=G_2$. We first rewrite $P_{V_w}(\mu,\Sig)$ as a composition of two function i.e., $P_{V_w}(\mu,\Sig) = P_w \circ P_V (\mu,\Sig)$. In particular, $P_V: \Re^d \times S_d^{++}(\Re) \to \Re^2$ i.e., $P_V(\mu,\Sig) = (\langle \mu,v\rangle,Trace(A\log \Sig))$ ($V=(v,A)$) and $P_w:\Re^2 \to \Re$ i.e.,  $P_w(x) = \langle w,  x\rangle$. We then can rewrite mixed Sliced Wasserstein distance as:
\begin{align*}
    \text{Mix-}SW_p^p(G_1,G_2) &=\mathbb{E}_{(w,v,A)\sim \mathcal{U}(\mathbb{S}) \otimes \mathcal{U}(\mathbb{S}^{d-1}) \otimes \mathcal{U}(S_d(\Re)) } [W_p^p(P_{V_w} \sharp G_1, P_{V_w}\sharp G_2)] \\
    &=\mathbb{E}_{(v,A)\sim \mathcal{U}(\mathbb{S}^{d-1}) \otimes \mathcal{U}(S_d(\Re)) } \left[ \mathbb{E}_{w \sim \mathcal{U}(\mathbb{S})}[W_p^p(P_w\sharp  P_{V} \sharp G_1, P_w \sharp P_{V}\sharp G_2)] \right] \\
    &= \mathbb{E}_{(v,A)\sim \mathcal{U}(\mathbb{S}^{d-1}) \otimes \mathcal{U}(S_d(\Re)) }  \left[SW_p^p( P_{V} \sharp G_1, P_{V} \sharp G_2 )\right].
\end{align*}
When $\text{Mix-}SW_p^p(G_1,G_2)=0$, it means that $SW_p^p( P_{V}
\sharp G_1, P_{V} \sharp G_2 )=0$ for $\mathcal{U}(\mathbb{S}^{d-1})
\otimes \mathcal{U}(S_d(\Re))$-almost every $(v,A)$. Using the
identity of indiscernibles proptery of
SW~\citep{bonnotte2013unidimensional}, we have $P_{V} \sharp G_1=
P_{V} \sharp G_2$ for $\mathcal{U}(\mathbb{S}^{d-1}) \otimes
\mathcal{U}(S_d(\Re))$-almost every $(v,A)$. Let denote
$\mathcal{F}[P_{V}\sharp G_1]$ and $\mathcal{F}[P_{V}\sharp G_2]$ as
the Fourier transform of $G_1$ and $G_2$ respectively, we have
$\mathcal{F}[P_{V}\sharp G_1]=\mathcal{F}[P_{V}\sharp G_2]$ for
$\mathcal{U}(\mathbb{S}^{d-1}) \otimes \mathcal{U}(S_d(\Re))$-almost
every $(v,A)$. Moreover, for all $y \in \Re^2$, we have: 
\begin{align*}
    \mathcal{F}[P_{V}\sharp G_1](y) &= \int_{\Re^2} e^{-2i\pi \langle y, x \rangle } d (P_{V}\sharp G_1)(x) \\
    % &= \int_{\Re^d \times S_d^{++}(\Re) }e^{-2i\pi \langle y, x \rangle}\mathrm{d} G_1(\mu_1,\Sig_1) \\
    &= \int_{\Re^d \times S_d^{++}(\Re) }e^{-2i\pi (y_1 \langle v,\mu_1\rangle + y_2 \langle A, \log \Sig_1\rangle_F )}\mathrm{d} G_1(\mu_1,\Sig_1)  \\
    &=\int_{\Re^d \times S_d^{++}(\Re) }e^{-2i\pi (\langle y_1  v,\mu_1\rangle +  \langle y_2 A, \log \Sig_1\rangle_F )}\mathrm{d} G_1(\mu_1,\Sig_1) \\
    &=\int_{\Re^d \times S_d^{++}(\Re) }e^{-2i\pi (\langle y_1  v,\mu_1\rangle +  \langle y_2 A, S_1\rangle_F )}\mathrm{d} ((Id,\log) \sharp G_1)(\mu_1,S_1) \\
    &= \mathcal{F}[(Id,\log) \sharp G_1](y_1 v,y_2 A).
\end{align*}
Therefore, we obtain $\mathcal{F}[(Id,\log) \sharp G_1](y_1 v,y_2 A) = \mathcal{F}[(Id,\log) \sharp G_2](y_1 v,y_2 A)$ for $\mathcal{U}(\mathbb{S}^{d-1}) \otimes \mathcal{U}(S_d(\Re))$-almost every $(v,A)$.  By injectivity of the Fourier transform, we get $(Id,\log) \sharp G_1 = (Id,\log) \sharp G_2$.  Since the function $f(\mu,\Sig) = (\mu,\log \Sig)$ is injective i.e., $f^{-1}(\mu,\Sig)=(\mu,\exp(\Sig))$, we obtain $G_1=G_2$, which completes the proof.

\subsection{Proof of Proposition~\ref{prop:boundness_smix}}
\label{subsec:proof:prop:boundness_smix}

 Since $P_{v,w}(\mu,\Sig) = w_1 \langle v,\mu \rangle+ w_2 \log(\sqrt{v^\top\Sig v})$ is a Borel meansurable, using Lemma 6 in ~\citep{paty2019subspace}, we have:
\begin{align*}
   & W_p^p(P_{v.w} \sharp G_1, P_{v,w}\sharp G_2)]  = \inf_{\pi_{v,w} \in \Pi(P_{v,w} \sharp G_1, P_{v,w}\sharp G_2)} \int_{\Re\times \Re} |x-y|^p \mathrm{d}\pi_{v,w}(x,y) \\
    &= \inf_{\pi \in \Pi( G_1, G_2)} \int_{\Re^d \times S_d^{++}(\Re) \times \Re^d \times S_d^{++}(\Re)} |P_{v,w}(\mu_1,\Sig_1)-P_{v,w}(\mu_2,\Sig_2)|^p \mathrm{d}\pi((\mu_1,\Sig_1),(\mu_2,\Sig_2)).
\end{align*}
Using the Minkowski's inequality, we have:
\begin{align*}
     W_p^p(P_{v.w} \sharp G_1, P_{v.w}\sharp G_2)] &\leq \inf_{\pi \in \Pi( G_1, G_2))} \int_{\Re^d \times S_d^{++}(\Re) \times \Re^d \times S_d^{++}(\Re)} 2^{p-1} \left(|P_{v.w}(\mu_1,\Sig_1)-P_{v.w}(\mu_0,\Sig_0)|^p\right. \\ &\quad \left.+|P_{v.w}(\mu_0,\Sig_0)-P_{v.w}(\mu_2,\Sig_2)|^p \right) \mathrm{d}\pi_{V_w}\mathrm{d}\pi((\mu_1,\Sig_1),(\mu_2,\Sig_2)) \\
    &= 2^{p-1} \left(\int_{\Re^d \times S_d^{++}(\Re)}|P_{v.w}(\mu_1,\Sig_1)-P_{v.w}(\mu_0,\Sig_0)|^p  \mathrm{d}G_1(\mu_1,\Sig_1)  \right.\\
    &\quad +\left. \int_{\Re^d \times S_d^{++}(\Re)}|P_{v.w}(\mu_0,\Sig_0)-P_{v.w}(\mu_2,\Sig_2)|^p  \mathrm{d}G_2(\mu_2,\Sig_2) \right).
\end{align*}
Moreover, from the Cauchy–Schwarz's inequality, we have:
\begin{align*}
    &|P_{v.w}(\mu_1,\Sig_1)-P_{v.w}(\mu_0,\Sig_0)| \\
    &=| w_1 \langle v,\mu_1 \rangle+ w_2 \log(\sqrt{v^\top\Sig_1 v})  -w_1 \langle v,\mu_0 \rangle+ w_2 \log(\sqrt{v^\top\Sig_0 v})| \\
    &=\left|w_1 \langle \mu_1-\mu_0,v\rangle  +0.5w_2 \log\left(\frac{v^\top\Sig_1 v}{v^\top\Sig_0 v}\right) \right| \\
    &\leq \sqrt{w_1^2+w_2^2} \sqrt{\langle \mu_1-\mu_0,v\rangle^2 +0.25 \log\left(\frac{v^\top\Sig_1 v}{v^\top\Sig_0 v}\right)^2 } \\
    &= \sqrt{\langle \mu_1-\mu_0,v\rangle^2 +0.25 \log\left(\frac{v^\top\Sig_1 v}{v^\top\Sig_0 v}\right)^2 } \\
    &\leq \sqrt{\|v\|_2^2 \|\mu_1-\mu_0\|_2^2 +0.25 \log\left(\max_{v}\frac{v^\top\Sig_1 v}{v^\top\Sig_0 v}\right)^2} \\
    &= \sqrt{ \|\mu_1-\mu_0\|_2^2 +  0.25\log(\lambda_{max}(\Sig_1,\Sig_2))^2 } =c((\mu_1,\Sig_0),(\mu_0,\Sig_0)),
\end{align*}
where $\lambda_{max}(\Sig_1,\Sig_0)$ is the  largest eigenvalue of the generalized problem $\Sig_1 v = \lambda \Sig_0 v$. From the assumption, we get:
\begin{align*}
    W_p^p(P_{v.w} \sharp G_1, P_{v,w}\sharp G_2)] 
    &\leq  2^{p-1} \left(\int_{\Re^d \times S_d^{++}(\Re)}c((\mu_1,\Sig_1),(\mu_0,\Sig_0))^p  \mathrm{d}G_1(\mu_1,\Sig_1)  \right.\\
    &\quad +\left. \int_{\Re^d \times S_d^{++}(\Re)}c((\mu_0,\Sig_0),(\mu_2,\Sig_2))^p \mathrm{d}G_2(\mu_2,\Sig_2) \right) < \infty,
\end{align*}
which completes the proof.

\subsection{Proof of Theorem~\ref{theorem:metricity_smixw}}
\label{subsec:proof:theorem:metricity_smixw}

\textbf{Symmetry, Non-negativity, and Triangle Inequality.} The symmetry, non-negativity, and triangle inequality of  SMix-W  can be obtained by following the proof for Mix-SW in Appendix~\ref{subsec:proof:theorem:mixSW_metric}. In this section, we focus  on the proof of identity of indiscernibles for SMix-W.

\paragraph{Identity of indiscernibles.} When $G_1=G_2$, we have
directly $\text{SMix-}W_p^p(G_1,G_2)=0$. We now prove that if
$\text{SMix-}W_p^p(G_1,G_2)=0$, we get $G_1=G_2$. From the definition
of SMix-W in Definition~\ref{def:smixW}, we have: 
\begin{align*}
    \text{SMix-}W_p^p(G_1,G_2) &=\mathbb{E}_{(w,v)\sim \mathcal{U}(\mathbb{S}) \otimes \mathcal{U}(\mathbb{S}^{d-1}) } [W_p^p(P_{v,w} \sharp G_1, P_{v,w}\sharp G_2)] \\
    &=\mathbb{E}_{v \sim \mathcal{U}(\mathbb{S}^{d-1})}[SW_p^p((Id,\log \circ \sqrt)\sharp P_v'\sharp G_1, (Id, \log \circ \sqrt)\sharp P_v'\sharp G_2)].
\end{align*}
When $\text{SMix-}W_p^p(G_1,G_2)=0$, it implies $SW_p^p((Id,\log \circ
\sqrt)\sharp P_v'\sharp G_1, (Id, \log \circ \sqrt)\sharp P_v'\sharp
G_2)=0$ for $\mathcal{U}(\mathbb{S}^{d-1})$-almost every $v$. Since
$\log(x)$ is an injective function, it leads to the fact that
$SW_p^p((Id, \sqrt)\sharp P_v'\sharp G_1, (Id,  \sqrt)\sharp
P_v'\sharp G_2)=0$ for $\mathcal{U}(\mathbb{S}^{d-1})$-almost every
$v$. By the identity of indiscernibles of
SW~\citep{bonnotte2013unidimensional}, we have $(Id,  \sqrt)\sharp
P_v'\sharp G_1=(Id,  \sqrt)\sharp P_v'\sharp G_2$ for
$\mathcal{U}(\mathbb{S}^{d-1})$-almost every $v$ with $P'_v(\mu,\Sig)
= (\langle v,\mu\rangle, v^\top \Sig v)$. Since the square root
function also injective on $\Re^+$, we have $P_v'\sharp G_1=
P_v'\sharp G_2$ which is equivalent to $P_v \sharp F_1 = P_v \sharp
F_2$ for $\mathcal{U}(\mathbb{S}^{d-1})$-almost every $v$ with
$P_v(x)=\langle v,x\rangle$ and $F_1=f*G_1$ and $F_2=f*G_2$ (f is the
Gaussian density kernel).  Let denote $\mathcal{F}[P_{v}\sharp F_1]$
and $\mathcal{F}[P_{v}\sharp F_2]$ as the Fourier transform of $F_1$
and $F_2$ respectively, we have $\mathcal{F}[P_{v}\sharp
F_1]=\mathcal{F}[P_{v}\sharp F_2]$ for
$\mathcal{U}(\mathbb{S}^{d-1})$-almost every $v$. Moreover, for all $t
\in \Re$, we have: 
\begin{align*}
    \mathcal{F}[P_{v}\sharp F_1](t) &= \int_{\Re^2} e^{-2i\pi t \epsilon } d (P_{v}\sharp F_1)(\epsilon) = \int_{\Re^d}e^{-2i\pi t\langle v,x \rangle }d F_1(x)  \\
&= \int_{\Re^d}e^{-2i\pi \langle tv,x \rangle }d F_1(x) = \mathcal{F}[F_1](tv).
\end{align*}
Therefore, we get $\mathcal{F}[F_1](tv) = \mathcal{F}[F_2](tv)$ for $\mathcal{U}(\mathbb{S}^{d-1})$-almost every $v$. By the injectivity of Fourier Transform, we get $F_1=F_2$ which leads to $G_1=G_2$ due to the identifiability of finite mixture of Gaussians (Proposition 2 in~\citep{yakowitz1968identifiability}), which concludes the proof.

 \begin{figure}[!t]
\begin{center}
  \begin{tabular}{c} 
\widgraph{1\textwidth}{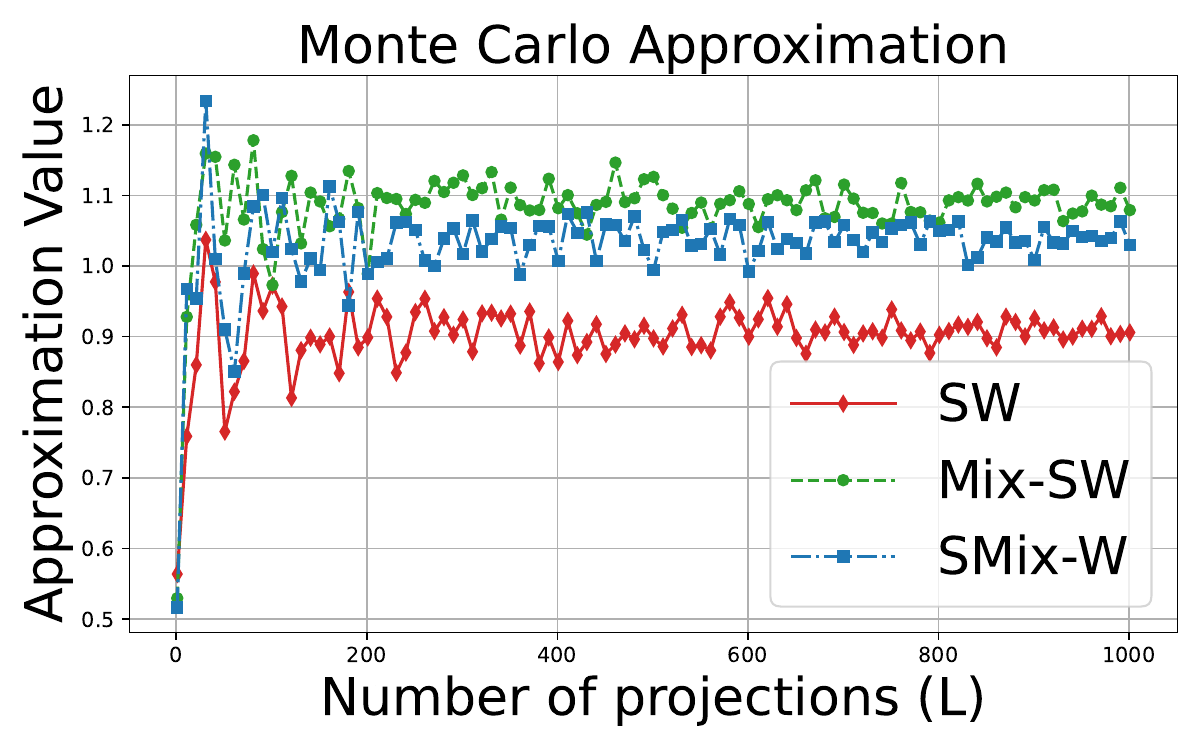} 

  \end{tabular}
  \end{center}
  \vspace{-0.3 in}
  \caption{
  \footnotesize{ {The figure show the approximation value of SW, Mix-SW, and SMix-W when increasing the number of projections $L$ (the number of Monte Carlo samples).}
}
} 
  \label{fig:MCapproximation}
\end{figure}

\subsection{{Monte Carlo Approximation}}
\label{subsec:MC_Approximation}
{
We investigate the Monte Carlo approximation of SW, Mix-SW, and SMix-W when increasing the number of projections $L$ (the number of Monte Carlo samples) in Figure~\ref{fig:MCapproximation}. In particular, we compare the mixing measures of the following two mixtures of Gaussians.
\begin{align*}
F_1(x) = & 0.3 \cdot \mathcal{N}\left(x;\, 
\begin{bmatrix} 0 \\ 0 \end{bmatrix},\ 
\begin{bmatrix} 1.0 & 0.8 \\ 0.8 & 1.5 \end{bmatrix}
\right)  +\ 0.4 \cdot \mathcal{N}\left(x;\, 
\begin{bmatrix} 3 \\ 3 \end{bmatrix},\ 
\begin{bmatrix} 0.7 & -0.3 \\ -0.3 & 0.9 \end{bmatrix}
\right) \\
& +\ 0.3 \cdot \mathcal{N}\left(x;\, 
\begin{bmatrix} -3 \\ 3 \end{bmatrix},\ 
\begin{bmatrix} 2.0 & 1.0 \\ 1.0 & 3.0 \end{bmatrix}
\right),
\end{align*}
and 
\begin{align*}
F_2(x) = & 0.2 \cdot \mathcal{N}\left(x;\, 
\begin{bmatrix} 1 \\ 1 \end{bmatrix},\ 
\begin{bmatrix} 1.2 & 0.5 \\ 0.5 & 0.8 \end{bmatrix}
\right)  +\ 0.5 \cdot \mathcal{N}\left(x;\, 
\begin{bmatrix} 4 \\ 4 \end{bmatrix},\ 
\begin{bmatrix} 0.4 & -0.2 \\ -0.2 & 0.6 \end{bmatrix}
\right) \\
& +\ 0.3 \cdot \mathcal{N}\left(x;\, 
\begin{bmatrix} -2 \\ 2 \end{bmatrix},\ 
\begin{bmatrix} 1.5 & 0.7 \\ 0.7 & 2.5 \end{bmatrix}
\right).
\end{align*}
We vary the number of projections (Monte Carlo samples) from $1$ to $1001$ with step size $10$. From Figure~\ref{fig:MCapproximation}, we see that the approximation values of SW, Mix-SW, and SMix-W converges well when increasing $L$. We observe that $L=100$ seems to give a good enough approximation for 2D mixtures of Gaussians. It is worth noting that we can adapt recent approaches such as control variates~\citep{nguyen2023control,leluc2024slicedwasserstein} and quasi Monte Carlo~\citep{nguyen2024quasimonte} to improve the standard convergence rate of Monte Carlo i.e., $\mathcal{O}(L^{-1/2})$. However, we believe that adapting such techniques is worth for a careful investigation. Therefore, we leave it to future works.
}
\end{document}